# Resistive switching suppression in metal/Nb:SrTiO$_3$ Schottky contacts prepared by room-temperature Pulsed Laser Deposition

R. Buzio[1,2*] and A. Gerbi[1,2]

[1]*CNR-SPIN Institute for Superconductivity, Innovative Materials and Devices, C.so F.M. Perrone 24, I-16152 Genova, Italy*

[2]*RAISE Ecosystem, Genova, Italy*

*Author to whom any correspondence should be addressed.
E-mail: renato.buzio@spin.cnr.it

**Abstract**

Deepening the understanding of interface-type Resistive Switching (RS) in metal/oxide heterojunctions is a key step for the development of high-performance memristors and Schottky rectifiers. In this study, we address the role of metallization technique by fabricating prototypical metal/Nb-doped SrTiO$_3$ (M/NSTO) Schottky contacts via Pulsed Laser Deposition (PLD). Ultrathin Pt and Au electrodes are deposited by PLD onto single-crystal (001)-terminated NSTO substrates and interfacial transport is characterized by conventional macroscale methods and nanoscale Ballistic Electron Emission Microscopy. We show that PLD metallization gives Schottky contacts with highly reversible current-voltage characteristics under cyclic polarization. Room-temperature transport is governed by thermionic emission with Schottky barrier height $\phi_B(\text{Pt/NSTO}) \sim 0.71 - 0.75 \text{eV}$, $\phi_B(\text{Au/NSTO}) \sim 0.70 - 0.83 \text{eV}$ and ideality factors as small as $n(\text{Pt/NSTO}) \sim 1.1$ and $n(\text{Au/NSTO}) \sim 1.6$. RS remains almost completely suppressed upon imposing broad variations of the Nb doping and of the external stimuli (polarization bias, working temperature, ambient air exposure). At the nanoscale, we find that both systems display high spatial homogeneity of $\phi_B$ ($\leq 50 \text{meV}$), which is only partially affected by the NSTO mixed termination ($|\phi_B(\text{M/SrO}) - \phi_B(\text{M/TiO}_2)| < 35 \text{meV}$). Experimental evidences and theoretical arguments - based on a metal-insulator-semiconductor description of the M/NSTO - indicate that the PLD metallization mitigates interfacial layer effects responsible for RS. This occurs thanks either to a reduction of the interfacial layer thickness and to the creation of an effective barrier against the permeation of ambient gas species affecting charge trapping and redox reactions. This description allows to rationalize interfacial aging effects, observed upon several-months-exposure to ambient air, in terms of a slow interfacial re-oxidation. Our work contributes to the fundamental understanding of interface-type RS and demonstrates that room-temperature PLD offers a viable platform for the realization of robust, RS-free NSTO-based Schottky contacts.

1. **INTRODUCTION**

Heterojunctions between high work function metals (e.g. Au, Pt, Cu, Ag...) and *n*-type NSTO single crystals attract great attention for their rich and yet not fully understood phenomenology. In fact, their response turns out to be very sensitive to the processing conditions so that different transport properties have been reported so far, even for the same nominal system. As a matter of fact, nearly-ideal M/NSTO Schottky Barrier Diodes (SBDs) with Schottky barrier height $\phi_B$ approaching the Schottky-Mott limit ($\phi_B$(Au/NSTO)~$\phi_B$(Pt/NSTO)~1.4eV, $\phi_B$(Cu/NSTO)~1.0eV) have been reported in a few studies, exploiting highly-oxidizing surface treatments of low-doped substrates [1–3]. On the other hand, reduced $\phi_B$ values (e.g. $\phi_B$(Au/NSTO)~0.78 − 0.88eV, $\phi_B$(Pt/NSTO)~0.70 − 0.97eV) ) together with the appearance of resistive switching (RS) have been documented by reports dealing with Room-Temperature (RT) metallization (e.g. thermal and *e*-beam evaporation, magnetron sputtering) of as-received or oxygen-annealed NSTO [4–10]. These SBDs show hysteretic current-voltage characteristics under cyclic polarization, reflecting bipolar resistance changes with anticlockwise polarity. Switching from a high-resistance to a low-resistance state occurs by applying a positive bias to the Schottky electrode, while the reverse switching takes place for the negative bias scan. The large variability in the response of M/NSTO SBDs is ascribed to extrinsic contamination and changes of the oxygen stoichiometry within an unintentional low-permittivity interfacial layer, formed in the NSTO near-surface region either during oxide surface preparation or metal evaporation. Theoretical and experimental studies indicate that the abundance of interfacial oxygen vacancies (i.e. a reduced interfacial state) decreases $\phi_B$ by several tens of an eV and enhances RS [11–13], while an optimal oxidation of the interface (oxidized state) has opposite effects [11,14–16]. There are evidences that for small current densities ($\ll 1 A/cm^2$), the resistance change occurs homogeneously over the interface without the need for an initial forming step, and it does not depend on the electrode material [17]. Besides reports describing interface-type RS with purely electronic effects (i.e. charge trapping/detrapping at oxygen-vacancy-induced defect states localized at the interfacial layer) [15,16,18–21], several studies invoke field-assisted oxygen-ions migration and interfacial redox reactions [17,22,23]. Importantly, switching devices are often found to interact with the environment, as both ambient oxygen and moisture may modulate $\phi_B$ and RS by promoting (or preventing) the excorporation/incorporation of ionic species within the interfacial layer [17,21,24–27]. Given the complex nature of the NSTO interfacial layer and of the RS mechanisms, attempts to improve reproducibility and uniformity of electronic transport across M/NSTO SBDs have moved along different directions including *in situ* NSTO surface cleaning [1,2,28], post-growth thermal treatments [14,29,30], insertion of oxide interlayers [17,19,30], modulation of ambient gases [21,24], and tailored growth of epitaxial electrodes [19]. Such investigations have practical relevance due to the crucial role played by M/NSTO SBDs in several proof-of-concept devices and applications (e.g. ferroelectric RAMs, memristors, spin injecting contacts, gas sensors) [24,31–33].

To deepen general understanding of the role of the metallization method, hereafter we explore SBDs prepared by Pulsed Laser Deposition (PLD) i.e. a versatile technique particularly suited to growth $SrTiO_3$-based heterostructures [34]. We show that RT PLD gives Pt/ and Au/NSTO SBDs characterized by a reduced $\phi_B$ ($\sim 0.7 - 0.8 eV$) and by the robust suppression of RS; this is indeed an interfacial condition rarely achieved by other physical vapour deposition techniques. We characterize transport properties of PLD-grown devices using a combination of macroscopic transport measurements and Ballistic Electron Emission Microscopy BEEM [35–38]. BEEM advantages compared to current-voltage or capacitive-voltage techniques are noteworthy, as it probes unbiased junctions and the evaluation of $\phi_B$ does not rely on data interpolation with specific transport models. Additionally, the temperature dependence of $\phi_B$ is readily obtained, together with fruitful knowledge of nanoscale spatial inhomogeneity and hot electrons transmittance across the target interfaces.

## 2. MATERIALS AND METHODS

The SBDs were prepared by depositing Au or Pt electrodes of ~10-15nm nominal thickness through a shadow mask (area $A \approx 2.3\text{mm}^2$) onto single-crystal NSTO (10×5×0.5 mm³ by PI-KEM LTD UK and CrysTec GmbH Germany; nominal doping $x_{Nb}$ = 0.01wt.%, 0.05wt.% and 0.5wt.%). Before metallization, we treated NSTO by a wet-etch method to obtain atomically flat (001)-terminated surfaces (see below). Deposition was accomplished in high vacuum ($\lesssim 5 \times 10^{-7}$mbar) using a KrF excimer laser (Lambda Physik) and high purity targets. Deposition conditions were: target-to-substrate distance 3cm, energy density on the target ~2 J cm$^{-2}$, repetition rate 10Hz, Pt deposition rate 0.5 nm/min., Au deposition rate $2.5 - 5.0$nm/min.. The Ohmic contacts were fabricated by PLD growth of a thick Al film on the backside of the substrates. After fabrication, SBDs were briefly exposed to ambient air for contact formation ($\approx 15$ minutes), next they were inserted into a Ultra High Vacuum UHV (~5×10$^{-9}$mbar) chamber for electronic transport characterization. In the following we refer to them as pristine devices. Since the response of M/NSTO SBDs can vary considerably upon exposure to different ambient gases [21,24,26,27], we also targeted variations of the transport properties due to exposure of the pristine devices to ambient air (relative humidity $\sim 40 - 60\%$) for a period of time going from a few hours up to several days and months. In line with [27], devices exposed to ambient air for a long time are referred to as aged devices.

Current-voltage (I-V) characteristics were measured with a two-probes method under dark (Figure 1(a)), from RT (T $\approx$ 293K) to liquid nitrogen temperature (T $\approx$ 78K). We used a Keithley 6430 sub-femtoamp sourcemeter, sweeping the voltage with a delay time of 2 s per 10 mV steps. We interpolated the forward-bias ($V > 0$) regions of the I-V curves with the thermionic-emission (TE) theory:

$$I = AA_R T^2 e^{-\phi_B^{I-V}/k_B T} e^{-qV/nk_B T} \qquad (1)$$

with the ideality factor $n$ and the effective Schottky barrier height $\phi_B^{I-V}$ as fitting parameters. The Richardson constant $A_R = 156 \text{ A cm}^{-2} \text{ K}^{-2}$ was used. To quantify RS effects, we focused on the hysteresis between the two forward-bias branches obtained by ramping the voltage up ($V\uparrow$) or down ($V\downarrow$) respectively, and we used the metrics based on the $\phi_B$ variation $\Delta\phi_{RS} \equiv |\phi_B^{I-V}(V\uparrow) - \phi_B^{I-V}(V\downarrow)|$ [4,19].

For selected SBDs, we also acquired (*ex situ*) RT capacitance *vs* voltage (C-V) characteristics with a Keithley 3322 LCZ meter. Modulation frequencies were in the range $1 - 100$KHz, whereas the test signal was 50mV.

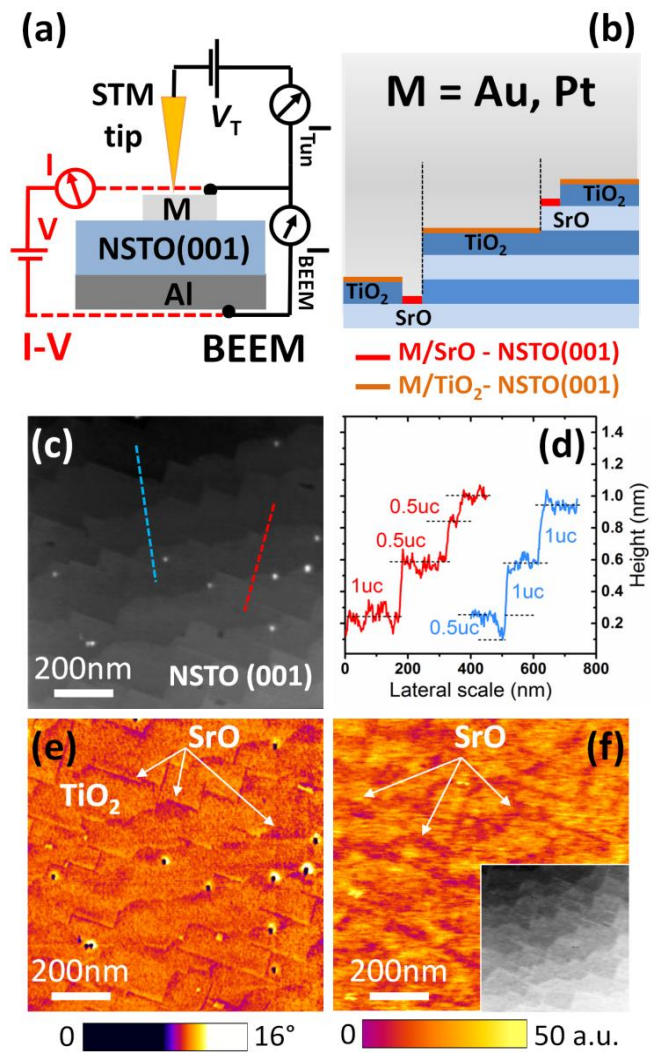

**Figure 1.** (a) Experimental set-up for I-V and BEEM measurements at M/NSTO SBDs. (b) Schematics (side-view) of a junction fabricated on a mixed-terminated substrate, showing $TiO_2$ terraces bordered by SrO domains (black dash lines mark the terrace ledges). (c) AM-AFM topography of wet-etched and annealed NSTO. (d) Cross section profiles corresponding to the dash lines in (c), reveal half unit-cell and one unit-cell steps between adjacent terraces. (e) Phase-lag map acquired simultaneously to (c), attesting peculiar contrast along the step edges as expected for a mixed-terminated surface. (f) FFM map (with the related topography in the inset), with domains of low friction arranged nearby the step edges.

BEEM was implemented under UHV with a commercial Scanning Tunneling Microscope STM (LT-STM by Omicron Nanotechnology GmbH Germany) equipped with an additional low-noise variable-gain current amplifier (custom DLPCA-200 by FEMTO GmbH Germany) [35,39,40]. Ballistic current $I_{BEEM}$ originates from hot electrons collected at the backside Ohmic contact, after travelling across the sample with kinetic energy $e|V_T|$ high enough to overcome the local energy barrier $\phi_B$ formed at the buried metal/NSTO interface (Figure 1(a)). We used negatively biased Au tips ($V_T < 0$), meaning that tunnelling electrons are injected from the tip to the Au electrode. For acquisition of BEEM spectra, tip voltage $V_T$ was ramped under feedback control, in this way keeping the tunnelling current ($I_{Tun} \sim 20 - 30$nA) constant. BEEM spectra were analyzed with custom software from National Instruments LabView. The BEEM maps were displayed using WSXM [41]. Besides investigating $\phi_B$ with nm spatial resolution and as a function of temperature T, we further explored the impact of the NSTO surface chemical termination thanks to the BEEM capability to probe (unbiased) metal/SrO − NSTO (M/SrO) and metal/$TiO_2$ − NSTO (M/$TiO_2$) interfaces fabricated simultaneously over the same substrate. In fact, the alternation of SrO and $TiO_2$ regions represents a naturally-occurring chemical termination stripe pattern, that can contribute to inhomogeneity and non-ideality of macroscopic transport. In this respect, we paid great care to characterize the bare surface morphology of the (001)-terminated NSTO substrates by *ex situ* Atomic Force Microscopy AFM (Solver P47H by NT-MDT), operated in ambient conditions either in dynamic mode (i.e. Amplitude-Modulation AM-AFM) or contact mode (i.e. Friction Force Microscopy FFM). Both AM-AFM and FFM are in fact able to distinguish the two chemical surface terminations [42,43]. We prepared atomically flat NSTO (001) surfaces following the "Arkansas" wet-etch method [44]. This recipe comprises: i) sinking NSTO in hot water to create soluble Sr-hydroxide complexes, ii) wet etching in Aqua Regia (3:1 vol. HCl–HNO$_3$) to selectively dissolve SrO at the NSTO surface, and iii) annealing in O$_2$ flow to induce surface reconstruction. Annealing was accomplished for 8 minutes at 1000°C or at 1150°C respectively, with 2-hr rump-up and 6-hr rump-down times in flowing O$_2$ (rate 120 $lh^{-1}$) [24,35]. The O$_2$ flux was maintained constant to improve the degree of surface oxygenation [24]. The "Arkansas" method is known to provide atomically flat NSTO with step-and-terrace surface structure and single TiO$_2$-termination [44]. However, mixed terminated surfaces can be obtained using this treatment [45], with minority SrO-regions coalescing and self-ordering along the step edges under the action of the thermally-activate surface diffusion induced by step iii) (Figure 1(b)) [42,46]. It is likely that such SrO domains originate either by incomplete removal of the Sr-hydroxides under non-optimized wet-etching, or by low-temperature ($\gtrsim 300$°C) Sr segregation as a result of the acidic "chemical cleavage" effect and/or thermal relaxation of the etched TiO$_2$ termination layer [47]. Representative results for a typical NSTO surface prepared in this study are displayed in Figure 1(c)-(f). The AM-AFM topography shows an atomically flat morphology with stepped-terrace structure (Figure 1(c)). Differently from the case of ideal surfaces with single TiO$_2$ termination – having straight and sharp step edges of 1 unit-cell (1uc~0.39nm) height [43] – here the terraces display both wavy and kinked ledges bordered by nanosized low-lying regions of 0.5uc height. Hence, both sharp 1.5uc steps and 0.5uc intermediate steps appear (Figure 1(d)) [48]. Occasionally, 1.5uc deep holes and 0.5uc height protrusions occur in the middle of terraces. All these regions can be associated to a local SrO termination. This assignment, besides being supported by topographical arguments, relies on the peculiar contrast displayed by AM-AFM and FFM maps along the step edges (Figure 1(e),(f)). The half uc domains are characterized by enhanced phase-lag and friction response than the rest of the surface, in qualitative agreement with previous evidences [42]. The surface SrO coverage – roughly estimated from phase lag maps *via* area thresholding criteria – stays in the range $\sim 11 - 16\%$. This is in line with estimates for incomplete $TiO_2$ termination ($\sim 88 \pm 2\,\mathrm{at.\%}$) previously found for annealed wet-etched NSTO surfaces [47].

For the electrostatic modelling of the metal/NSTO junctions, we implemented with Wolfram Mathematica a Metal-Insulator-Semiconductor (MIS) one-dimensional model, with a temperature- and field-dependent permittivity $\epsilon_s(E,T)$ for NSTO, and a constant thickness $\delta_i$ and dielectric constant $\epsilon_i$ for the insulating layer. The MIS model predicts a strong dependence of $\phi_B$ on the applied bias, temperature, Nb doping and interfacial layer thickness (e.g. see Shimizu *et al.*[49] and Kamerbeek *et al.*[50]; for further details see the Supplementary Information Section S6 in [36]).

## 3. RESULTS AND DISCUSSION

*3.1 Macroscopic transport properties of the PLD-grown Schottky contacts*

Macroscopic RT I-V measurements showed an excellent rectifying behaviour for PLD-grown Pt/NSTO SBDs, with rectification ratio $\approx 2 \times 10^4$ at $\pm 0.4\mathrm{V}$, Schottky barrier height $\phi_B^{I-V} \cong 0.75\mathrm{eV}$ and ideality factor $n \cong 1.10$ at the lowest doping $x_{\mathrm{Nb}} = 0.01\mathrm{wt.\%}$ (Figure 2(a)). The barrier height $\phi_B^{I-V} \cong 0.75\mathrm{eV}$ is in line with $\phi_B^{C-V} \sim 0.71 - 0.77\mathrm{eV}$ estimated from RT C-V measurements (Supplementary Information section S1), also it agrees with a number of studies preparing Pt/NSTO junctions at RT by RF and DC sputtering [7,8] or *e*-beam evaporation [4,6]. The ideality factor $n = 1.10$ is close to the ideal case ($n = 1$) and fits the lowest values reported for NSTO single-crystals, namely $n = 1.14$ for low-doped ozone-cleaned Au/NSTO junctions [49] and $n = 1.19$ for highly-doped all-epitaxial Pt/NSTO interfaces [19]. Inspection of I-V-T measurements (Figure 2(b)) showed that the junction response depends on temperature T, as a simultaneous lowering of $\phi_B^{I-V} \to 0.26$eV and increase of $n \to 3.8$ took place on decreasing T from RT to 78K. This is expected in the framework of the MIS description of the system where the T-dependence arises from both the temperature-dependent permittivity of NSTO, and the presence of a low-permittivity interfacial layer [36,51]. Given the good linear correlation of the $\phi_B^{I-V}$ vs $n^{-1}$ data at $x_{\text{Nb}} = 0.01$wt.% (Figure 2(c)) [50,51], we estimated the fundamental flat-band barrier height $\phi_{FB} \equiv eV_{FB}$ and semiconductor degeneracy $\xi_F$ by interpolation with the equation [52]:

$$\phi_B^{I-V} = (\phi_{FB} - \xi_F)n^{-1} + \xi_F \qquad (2)$$

thus obtaining $\phi_{FB} \approx 0.94$eV and $\xi_F \approx 40$meV. The positive value of $\xi_F$ is expected for the non-degenerate NSTO $(20 - 200\text{meV})$ [51,53], whereas $\phi_{FB}$ represents a temperature- and electric-field-invariant property of the PLD-grown Pt/NSTO interface [51,53]. For $x_{\text{Nb}} = 0.5$ wt.%, we obtained similar RT I-V characteristics with rectifying ration $\approx 50$ at $\pm 0.35$V and $\phi_B^{I-V} \cong 0.71$eV (Supplementary Information S2). Here, the higher doping increased both the reverse-bias leakage and the ideality factor ($n \approx 1.55$), likely due to enhanced thermally-assisted tunnelling contributions and/or conduction through the interface states [9,54]. Moreover, we observed poor linear correlation of the $\phi_B^{I-V}$ vs $n^{-1}$ data over the range $78\text{K} - 293\text{K}$. This indicates that the interface energetics was not fully rationalized by Eq. (2) for $x_{\text{Nb}} = 0.5$ wt.%.

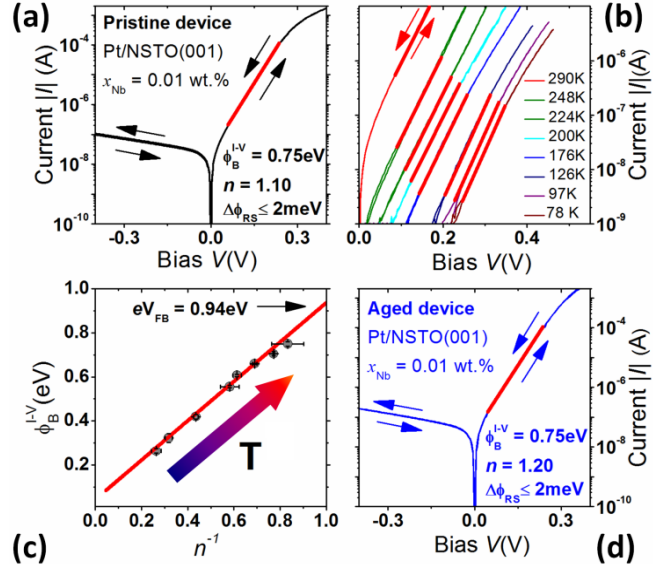

**Figure 2.** a) Reversible I–V curve for a pristine PLD-grown Pt/NSTO SBD, measured at RT in UHV under cyclic polarization. Sweep directions are indicated by arrows. Interpolation with TE is shown (red curve). b) Temperature-dependent I–V curves (forward branch) for the device in (a). c) The $\phi_B^{I-V}\ vs\ n^{-1}$ plot obtained from I-V-T curves. Linear interpolation (red curve) gives the flat-band barrier height $eV_{FB} = 0.94\text{eV}$. (d) The I-V characteristic of the Pt/NSTO SBD upon 60 months aging in ambient air.

Whereas all the PLD-grown SBDs had TE parameters comparable with those reported in literature, their most notable property was that hysteresis related to interfacial-type RS was almost completely suppressed in the RT forward bias region. For $x_{\text{Nb}} = 0.01\text{wt.}\%$ we found $\Delta\phi_{RS} \leq 2\text{meV}$, with voltage loop amplitudes going from $\pm 0.4\text{V}$ up to a few Volts (Supplementary Information S3). Table 1 reveals that such a small value for $\Delta\phi_{RS}$ is rare for Pt/NSTO interfaces processed at RT. For high-temperature processing, suppressed RS with $\Delta\phi_{RS}\sim 40\text{meV}$ has been claimed so far only for all-epitaxial Pt(001) electrodes deposited at 800°C. Table 1 also shows that the RS suppression persisted for the higher doping $x_{\text{Nb}} = 0.5\text{wt.}\%$ ($\Delta\phi_{RS} \leq 4\text{meV}$). Importantly RS remained suppressed when cooling PLD-grown Pt/NSTO SBDs to 78K. Also, the I-V characteristics were unaffected by variations of the oxygen pressure and moisture in the working environment, specifically imposed by moving the pristine devices from vacuum to laboratory air for a given period of time. Contrary to previous studies [24,27], we did not observe variations of the TE parameters upon exposure of the PLD-grown Pt/NSTO devices to ambient air for a few hours up to several days. We also studied aging of the Pt-based SBDs upon 60 months exposure to ambient air: in such case we just noticed small variations of $\phi_B^{I-V}$ and/or of $n$, but without any enhancement of the RS hysteresis (Supplementary Information S2).

| $x_{Nb}$ (wt.%) | Prep. method | $\phi_B^{I-V}$ (eV) | n | $\Delta\phi_{RS}\downarrow$ (eV) | Ref. |
|---|---|---|---|---|---|
| 0.7 | e-beam | 0.97 | 1.80 | 0.46 | [19] |
| 0.5 | DC-sputt | 1.2 | 1.5 | 0.35 | [10] |
| 0.7 | DC-sputt | 0.94 | 1.40 | 0.17-0.30 | [19] |
| 0.1 | DC-sputt | 1.24 | 1.7 | 0.12 | [22] |
| 0.1 | e-beam | 0.77 | 1.40 | 0.11 | [6] |
| 0.1 | e-beam | 0.77 | 1.60 | 0.10 | [4] |
| 0.05 | RF-sputt | 0.78 | 1.40 | 0.10 | [7] |
| 0.5 | DC-sputt | 0.64* | 7.2* | 0.03* | [8] |
| 0.01 | PLD | 0.75 | 1.10 | ≤0.002 | This study |
| 0.5 | PLD | 0.71 | 1.55 | ≤0.004 | This study |

**Table 1.** Transport parameters for Pt/NSTO SBDs prepared on single-crystal NSTO by RT metallization. The $\phi_B^{I-V}$ and $n$ values derive from fitting the high-resistance branch with the TE model whenever the RT I-V characteristics display bipolar RS. Symbol * refers to parameters we estimated from digitized published plots.

For the PLD-grown Au/NSTO SBDs, macroscopic RT I-V measurements showed good rectification with rectification ratio $> 10^3$ at $\pm 0.5$V for $x_{Nb} = 0.01$wt.% (Figure 3(a)). Compared to Pt/NSTO devices, the reverse-bias current was higher ($\sim 2 \times 10^{-7}$A at $-0.3$V) and a non-ideal double slope structure emerged under forward bias. This corresponds to the higher barrier $\sim 0.83$eV (with smaller ideality factor $n \sim 1.6$) at the high bias $V > 0.25$V and a smaller barrier $\sim 0.78$eV (with ideality factor $n \sim 2.1$) at the low bias $0 < V < 0.25$V. As we observed this feature for all the prepared Au/NSTO devices, it is likely that the Au metallization by PLD led to a patched interface [55] with the low-barrier patches dominating at small bias. Both barriers 0.78eV, 0.83eV agree with previous studies preparing Au electrodes by RT magnetron sputtering [5] or by *e*-beam evaporation [4,9]. For the PLD-grown interface, the variation of electronic transport with T was remarkable. Focusing on the high-barrier region (Figure 3(b)), lowering of $\phi_B^{I-V} \sim 0.83\text{eV} \rightarrow 0.23\text{eV}$ and increase of $n \sim 1.6 \rightarrow 6.9$ took place on decreasing T from RT to 78K. Interpolation with Eq. (3) provided $\phi_{FB} \approx 1.26\text{eV}$ and $\xi_F \approx 70\text{meV}$ (Figure 3(c)). Here, we ascribe the higher $\xi_F$ value to the non-ideality of the forward I-V branches, that implies a less accurate determination of $\phi_{FB}$ and $n$.

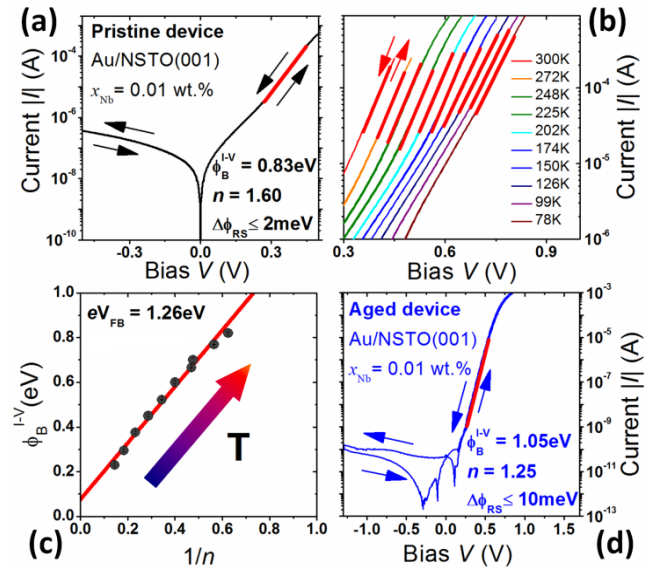

**Figure 3.** a) Reversible I–V curve of a PLD-grown Au/NSTO SBD, measured at RT in UHV under cyclic polarization. Sweep directions are indicated by arrows. The non-ideal two barrier structure demands interpolation with TE emission in two separate regions. b) Temperature-dependent I–V curves (forward branch, high-barrier region) for the device in (a). c) The $\phi_B^{I-V}$ vs $n^{-1}$ plot obtained from temperature-dependent I-V curves in b). Linear interpolation (red curve) gives an estimate of the fundamental flat-band barrier height $eV_{\text{FB}} = 1.26\text{eV}$. (d) Current-voltage I-V characteristic of the aged device in (a).

Also for Au/NSTO SBDs, RS hysteresis was suppressed in the forward bias region with $\Delta\phi_{RS}$

amounting to a few meV regardless of the NSTO doping ($x_{Nb} = 0.01\text{wt.}\%, \; 0.05\text{wt.}\%$) and of the working temperature ($78K - 293K$, Supplementary Information S4). Again, according to Table 2, such a small $\Delta\phi_{RS}$ is rare for Au/NSTO interfaces processed at RT.

| $x_{Nb}$ (wt.%) | Prep. method | $\phi_B^{I-V}$ (eV) | n | $\Delta\phi_{RS}$ ↓ (eV) | Ref. |
|---|---|---|---|---|---|
| 1.4 | e-beam (140°C) | 1.48 | 1.8 | 0.83 | [21] |
| 0.5 | therm | 1.06 | 2.20 | 0.27 | [36] |
| 0.05 | therm | 1.33 | 1.60 | 0.20 | [24] |
| 0.05 | sputt | 0.88 | 2.0 | 0.12 | [5] |
| 0.01 | e-beam | 1.15 | 1.1 | 0.07* | [9] |
| 0.05 | sputt | 0.81* | 2.0* | 0.06* | [5] |
| 0.1 | e-beam | 0.78 | 2.0 | 0.06* | [9] |
| 0.01 | therm | 1.30 | 1.30 | 0.05 | [36] |
| 0.1 | e-beam | 0.88 | 2.16 | 0.04 | [4] |
| 0.05 | PLD | 0.70 | 2.20 | ≤0.007 | This study |
| 0.01 | PLD | 0.83 | 1.60 | ≤0.004 | This study |

**Table 2.** Transport parameters for Au/NSTO SBDs prepared on single-crystal NSTO by RT metallization. The $\phi_B^{I-V}$ and $n$ values derive from fitting the high-resistance branch with the TE model whenever the RT I-V characteristics display bipolar RS. Symbol * refers to parameters we estimated from digitized published plots.

Contrary to the PLD-grown Pt/NSTO devices, the Au/NSTO ones showed aging in ambient air. As shown in Figure 3(d), a 60-months-aging removed the non-ideal double-slope structure of the I-V characteristics, leading to an exponential rise of the current with rectification ratio $> 10^7$ at $\pm 1V$, $\phi_B^{I-V} \approx 1.05\text{eV}$ and improved ideality factor $n \sim 1.25$ (see Supplementary Information S4). The RS hysteresis in the positive bias branch was almost unaffected by aging, with $\Delta\phi_{RS} \leq 10\text{meV}$, whereas a small hysteresis appeared for negative biases and small currents $|I| < 100\text{pA}$. This effect very likely relates to the oxygen diffusion through the Au electrode and involves the effective reoxygenation of oxygen-related defects (e.g. oxygen vacancies) induced by the PLD metallization (see below). The process is thought to increase $\phi_B^{I-V}$ and to improve rectifying quality in oxygen-rich atmospheres by suppressing the reverse leak current [24]. Albeit we did not explore in detail the timescale of the aging process, this is certainly much longer than previously reported for a similar system. In fact, Buzio et al.[24] showed that ambient oxygen reversibly modulates the response of the thermally-evaporated Au/NSTO SBDs over a timescale going from several minutes to a few hours, whereas this was not observed when exposing the PLD-grown pristine devices to air or vacuum respectively. Also, the interfacial aging embodied by Figure 3(d) differs qualitatively from the aging phenomenology discussed for the magnetron-sputtered Au/NSTO interface by Hirose et al.[27]. Here, they found that the Au/NSTO junctions - stored in vacuum or in a purified Ar atmosphere - showed small rectification and negligible RS hysteresis immediately after their fabrication, whereas enhanced rectification (with $\phi_B^{I-V} \sim 0.65 - 0.85$eV) and large RS hysteresis appeared upon a-few-days exposure to the oxygen-rich ambient air.

*3.2 Nanoscale transport properties of the PLD-grown Schottky contacts*

Figure 4(a),(b) shows the STM topography and BEEM maps acquired over a representative region of a PLD-grown ultrathin Au electrode (deposited on the NSTO substrate of Figure 1(c)-(f)). The topography (Figure 4(a)) is typical of a polycrystalline metal overlayer and appears homogeneous. At higher magnification the surface displays nanosized atomically-smooth facets that represent the characteristic morphologies of a continuous film regime [56] (Supplementary Information S5). The surface heights' range ($<$ 5nm) is smaller than the nominal thickness ($\sim$15nm), which supports the picture of a continuous and conformal substrate coating. Specifically, there is no evidence of narrow trenches that arrive down to the substrate, as found for thermally-evaporated Au pads affected by liquid-like metal clustering [36,37]. The BEEM map (Figure 4(b),(d)) reveals a peculiar spatial modulation of the hot-electrons current $I_{BEEM}$ injected across the buried Au/NSTO interface. The spatial anisotropy of the BEEM pattern indicates a tight correlation of $I_{BEEM}$ with the local surface structure of NSTO. Given the large oscillations of the Au morphology (~1nm) compared to the NSTO surface steps, terraces cannot be distinguished by visual inspection of the STM topography and correlated to the BEEM pattern. Nonetheless, a statistical comparison of the Au morphology and BEEM signal through autocorrelation functions indicates that the two share a common spatial periodicity (Figure 4(e)), with the same wave vector and wavelength ~120nm. This value corresponds well to the width of NSTO terraces (Figure 1(e),(f)), hence to the NSTO surface structure. Thanks to the very intense contrast between low- and high-transmitting domains on the BEEM map (× 6, Figure 4(d)), one can appreciate their kinked and wavy borders that closely resemble the geometry of the NSTO terrace ledges and minority SrO domains. We thus conclude that the BEEM pattern reflects specific contributions from the $TiO_2$ and SrO terminations. We associate the high-transmitting domains to a local Au/$TiO_2$ interface, and the low-transmitting regions to a local Au/SrO junction (Supplementary Information S6).

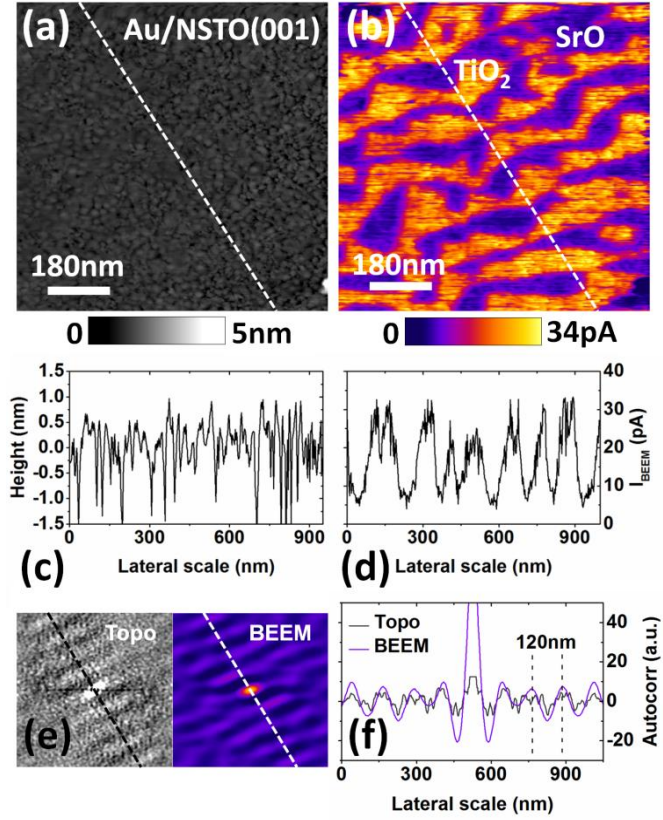

**Figure 4.** a) STM topography and b) BEEM map acquired simultaneously over a representative Au region of a PLD-grown Au/NSTO(001) SBD ($I_T = 20$nA, $V_T = -1.95$V, $T = 80$K, $x_{Nb} = 0.01$wt.%). BEEM reveals strong spatial modulation associated to the mixed surface termination of NSTO. (c) Cross-section height along the dashed lined in (a). (d) Cross-section ballistic current along the dashed line in (b). (e) Autocorrelations of the topography and BEEM maps in (a),(b). (f) Cross-sections along the dashed lines in (e).

In Figure 5(a)-(d) we report similar results for a representative PLD-grown Pt/NSTO SBD, prepared on a mixed-terminated NSTO substrate by depositing a 10nm-thick electrode.

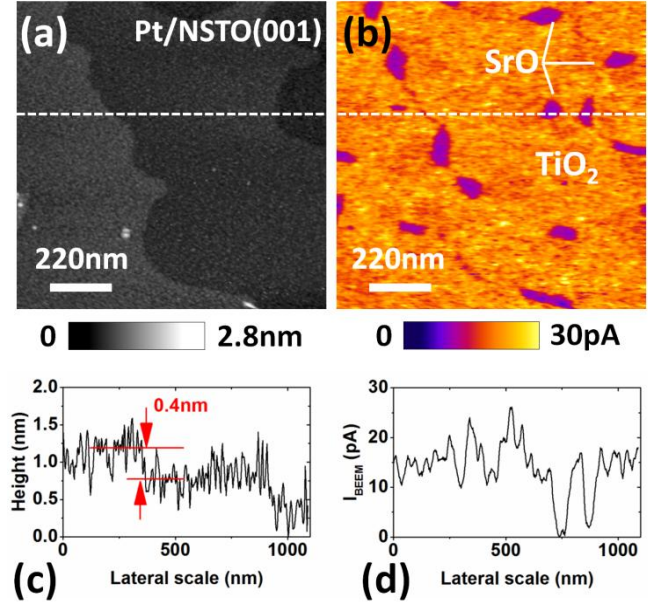

**Figure 5.** a) STM topography and b) BEEM map acquired simultaneously over a representative Pt region of a PLD-grown Pt/NSTO SBD ($I_T = 22$nA, $V_T = -1.8$V, $T = 272$K; $x_{Nb} = 0.05$wt.%). (c) Cross-section height along the dashed lined in (a), attesting the small roughness of the Pt electrode and conformal coating of the stepped NSTO substrate. (d) Cross-section ballistic current along the dashed line in (b).

The topography (Figure 5(a)) reveals a granular structure, with grains of ~10nm lateral size that conformally coat the stepped NSTO substrate. As shown in the related cross-section (Figure 5(c)), the range of surface heights over atomic terraces (~0.5nm) is small with respect to the nominal film thickness and turns out to be comparable with 1uc of NSTO. Together with the tight packing of the small grains, this results into a very compact Pt overlayer. The nanocrystalline morphology of the electrode qualitatively agrees with that found for other PLD-grown oxide-based SBDs, e.g. Pt/(100)$\beta - Ga_2O_3$ [37]. The associated BEEM map (Figure 5(b)) shows that ballistic injection is homogeneous across the interface, albeit a strong decrease of $I_{BEEM}$ (by ~70% to ~90%) occurs over well-defined nanometric spots (Figure 5(d)). This fact, together with their arrangement along the surface step edges, indicates their tight correspondence with localized Pt/SrO junctions, whereas the remaining part of the BEEM map can be assigned to the Pt/$TiO_2$ interface.

Qualitatively, this agrees with the BEEM contrast at PLD-grown Au/NSTO SBDs.

To quantify the Schottky barrier height fluctuations induced by the NSTO mixed termination, we compared BEEM spectra acquired respectively across M/TiO$_2$ and M/SrO junctions. In Figure 6(a) we consider the case of two representative average spectra corresponding to Au/TiO$_2$ and Au/SrO interfaces at T = 80K (Supplementary Information S7). Each spectrum shows a monotonic behaviour, with the characteristic threshold V$_{th,SB}$ corresponding to the local value of the barrier height $\phi_{B0} = e|V_{th,SB}|$. As expected, the two spectra show that the intensity of the current I$_{BEEM}$ transmitted across the Au/TiO$_2$ interface is higher than at the Au/SrO one. By interpolation with the Ludeke and Prietsch (LP) model, i.e. $I_{BEEM}/I_{Tun} = R(V_T - |V_{th,SB}|)^{5/2}$ (interpolation range $0.4V - 1.2V$), we estimated both $\phi_{B0}$ and the ballistic transmittance $R$. This indicated that the main effect of the NSTO surface termination is to affect the $R$ value, from $R \sim 1.6 \times 10^{-4} eV^{-5/2}$ for the Au/SrO interface to $R \sim 3.8 \times 10^{-4} eV^{-5/2}$ for the Au/TiO$_2$ one, whereas the two average Schottky barrier heights differ only by a few tens of meV, i.e. $\phi_{B0}(Au/TiO_2) - \phi_{B0}(Au/SrO) \sim 28 meV$. By considering a few surface spots over the same device we found that the difference between the two barriers was positive (($\phi_{B0}(Au/TiO_2) > \phi_{B0}(Au/SrO)$)) and varied in the range $\sim 5 meV - 36 meV$ (Supplementary Information S8). In Figure 6(b) we report the dual parameters distributions and the $\phi_{B0}$ histograms obtained by interpolating with the LP model a large ensemble of individual spectra (about 5700) acquired randomly, at 80K, over the Au electrode. Local variations in the intensity and onsets of such spectra caused a remarkable spreading of the local $\phi_{B0}$ and R parameters, in agreement with the current fluctuations already observed in the BEEM map (Figure 4(b)). The barrier spread is from $\sim 0.65 eV$ up to $\sim 0.95 eV$, and ballistic transmittance $R$ varies from $\sim 1 \times 10^{-4} eV^{-5/2}$ up to $\sim 2 \times 10^{-3} eV^{-5/2}$. The associated histogram of barrier heights is centered at the ensemble-averaged value $\overline{\phi}_{B0} \approx 0.80 eV$. The statistical spread of the histogram originates from two distinct contributions, namely the spatial variations of the Schottky barrier height and the measurement noise [57]. Careful analysis indicates that the experimental uncertainty contributes to the spread by (at most) ~30meV (Supplementary Information S9), so that the actual barrier inhomogeneity at the PLD-grown Au/NSTO SBDs can be evaluated to be of about $\sigma \approx 50$meV (standard deviation).

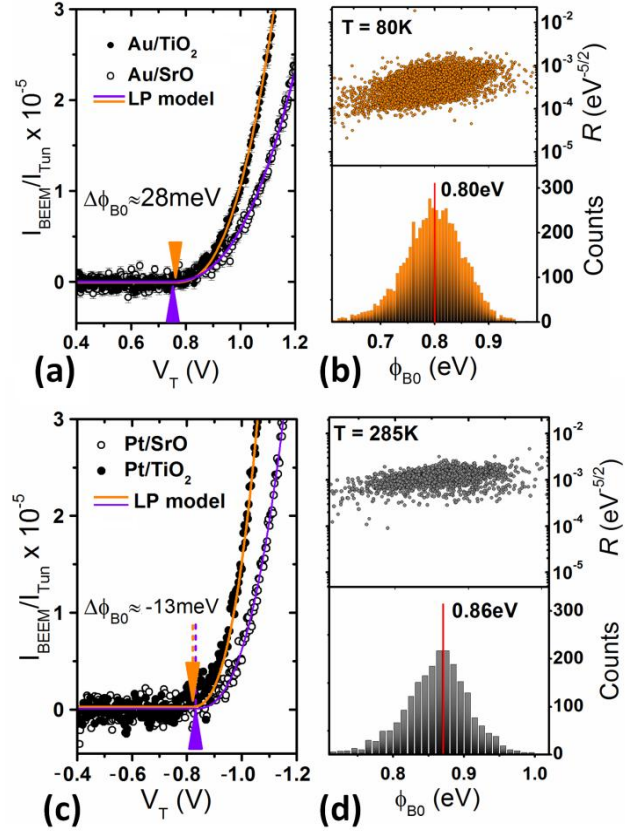

**Figure 6.** (a) Average BEEM spectra for PLD-grown Au/NSTO SBD, acquired on selected regions associated to different surface terminations of NSTO ($x_{Nb} = 0.01$wt.%, T= 80K). (b) Dual parameters ($\phi_{B0}, R$) distribution (top) and $\phi_{B0}$ histogram (bottom) for the Au/NSTO junction, obtained by fitting an ensemble of ~5700 spectra with the LP model. (c) Average BEEM spectra for PLD-grown Pt/NSTO SBD, showing the effect of different terminations of NSTO ($x_{Nb} = 0.01$wt.%, T= 285K). (d) Dual parameters ($\phi_{B0}, R$) distribution (top) and $\phi_{B0}$ histogram (bottom) for the Pt/NSTO junction (from an ensemble of ~2200 spectra).

Figure 6(c) displays the role of the mixed NSTO surface termination from average BEEM spectra acquired at specific locations of PLD-grown Pt/NSTO SBD (Supplementary Information S10). Also in this case, the main effect of the NSTO surface termination was to impact the $R$ value roughly by a factor $\sim 2$, from $R \sim 5.0 \times 10^{-4} \text{eV}^{-5/2}$ for the Pt/SrO interface to $R \sim 1.1 \times 10^{-3} \text{eV}^{-5/2}$ for the Au/TiO$_2$ one, whereas the two average barrier heights differ by a few meV, i.e. $\phi_{B0}(\text{Pt/TiO}_2) - \phi_{B0}(\text{Pt/SrO}) \sim -13 \text{meV}$. By considering a few surface spots over the same device we found that the difference between the two barriers was in this case negative (($\phi_{B0}(\text{Pt/TiO}_2) < \phi_{B0}(\text{Pt/SrO})$) and varied in the range from $-13 \text{meV}$ to $-35 \text{meV}$ (Supplementary Information S11). Whereas the origin for the sign inversion deserves further investigations, the small difference $|\phi_{B0}(\text{M/TiO}_2) - \phi_{B0}(\text{M/SrO})| \leq 35 \text{meV}$ turns out to be in reasonable agreement with the work function differences detected on mixed-terminated single-crystal SrTiO$_3$ substrates using Kelvin-Probe AFM and Low-Energy Electron Microscopy LEEM ($\sim 10 - 70 \text{meV}$) [45,58,59]. In Figure 6(d) we also report the dual parameters distributions and the $\phi_{B0}$ histograms obtained by interpolating with the LP model a large ensemble of individual spectra (about 2200) acquired randomly, at 285K, over the Pt electrode. The barrier spread is from $\sim 0.77 \text{eV}$ up to $\sim 0.97 \text{eV}$, and ballistic transmittance $R$ varies from $\sim 3 \times 10^{-4} \text{eV}^{-5/2}$ up to $\sim 3 \times 10^{-3} \text{eV}^{-5/2}$. The associated histogram of barrier heights is centered at the ensemble-averaged value $\overline{\phi}_{B0} \approx 0.86 \text{eV}$, whereas the statistical spread can be now evaluated to be $\sigma \sim 55 \text{meV}$ (standard deviation). The most striking result is that either Au/NSTO and Pt/NSTO junctions have a nanoscale $\phi_{B0}$ spread which is half of that measured by BEEM for thermally-evaporated SBDs ($\sim 100 \text{meV}$) [36], which is indicative of an enhanced electrostatic homogeneity of the PLD-grown interfaces.

In analogy with our previous BEEM studies of thermally-evaporated junctions [36], we now focus on the T-dependence of $\overline{\phi}_{B0}$ for low-doped PLD-grown SBDs. In fact, we showed that for $x_{\text{Nb}} = 0.01 \text{ wt.\%}$, $\overline{\phi}_{B0}$ $vs$ T graphs allow to estimate the interfacial layer capacitance $C_i$ and the flat-band barrier height $eV_{\text{FB}}$ within a MIS electrostatic analysis [50]. The T-dependence of $\overline{\phi}_{B0}$ originates from two concurrent factors, namely the temperature-dependent permittivity of NSTO and the presence of a low-permittivity interfacial layer [35,49] making the junction to behave as a MIS heterostructure. This ultimately results into a monotonic dependence of $\overline{\phi}_{B0}$ on T, due to a progressive redistribution of the flat-band voltage $eV_{FB}$ between the NSTO depletion layer and the interfacial layer, driven by T-variations of the depletion width. A schematics of the temperature evolution of band bending at the insulator/NSTO interface is reported in Figure 7(a). Figure 7(b) attests that $\bar{\phi}_{B0}$ varies by ~200meV for the Au/NSTO interface over the range from RT to 78K, whereas a decrease of only ~50meV takes place for the Pt/NSTO junction. This SBH variations are smaller than those estimated from macroscale I-V measurements ($\phi_B^{I-V}$ $vs$ T plots in Figures 2(c), 3(c)). Indeed, as BEEM probes unbiased junctions, the $\bar{\phi}_{B0}$ $vs$ T plots are not affected by transport modeling.

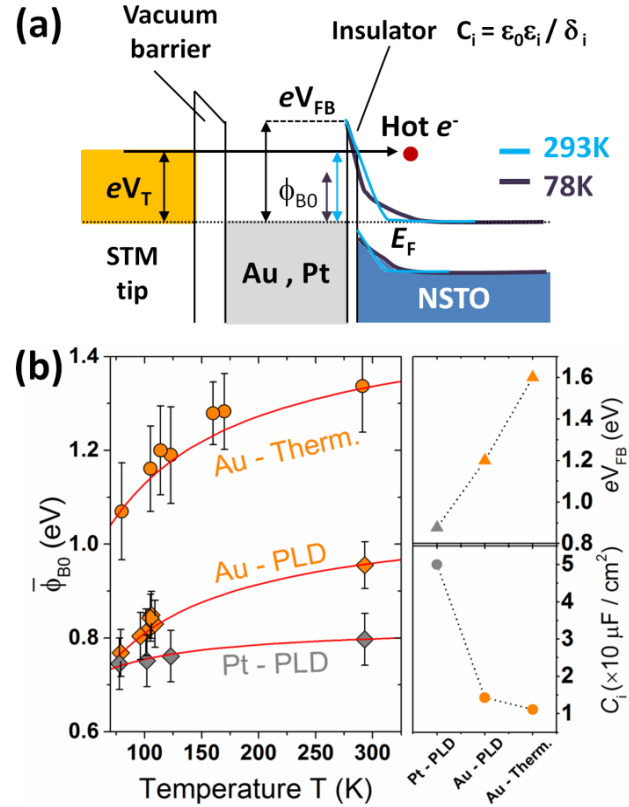

**Figure 7.** a) Schematics of energy band diagram for an unbiased junction (not in scale). Hot electrons emitted by the STM tip with high enough kinetic energy, tunnel across the thin insulator layer into the conduction band of NSTO and generate a ballistic current $I_{BEEM}$. The onset of ballistic current $\phi_{B0}$ varies according to the temperature evolution of the NSTO surface potential at the insulator/NSTO interface. b) (left panel) $\bar{\phi}_{B0}$ $vs$ T for pristine PLD-grown SBDs ($x_{Nb} = 0.01$ wt.%): error bars correspond to $\pm 1\sigma$. BEEM data for thermally-evaporated Au/NSTO diodes are shown for comparison (from [36]). Solid lines are the MIS model predictions, calculated for specific combinations of eV$_{FB}$, $C_i$ parameters (see right panels).

In this framework, a MIS electrostatic analysis [50] of the metal/insulator/NSTO heterostructure reasonably agrees with the BEEM results, provided that one treats the dielectric permittivity of NSTO through the phenomenological equation [53] $\epsilon_s(E,T) = b(T)/\sqrt{a(T) + E^2}$ ($E$ electric field, $a$ and $b$ material parameters), and introduces an electrode-dependent interfacial layer capacitance ($C_i \approx 14\mu F/cm^2$ for Au/NSTO; $C_i \approx 50\mu F/cm^2$ for Pt/NSTO) and the flat-band barrier height $eV_{FB}$ ($\approx 1.13eV$ for Au/NSTO; $\approx 0.88eV$ for Pt/NSTO) (Supplementary Information S12). Estimates for the flat-band barrier height reasonably agree with those obtained from I-V-T analysis ($\approx 1.26eV$ for Au/NSTO; $\approx 0.94eV$ for Pt/NSTO). Comparison with the thermally-evaporated Au/NSTO SBDs ($C_i \approx 11\mu F/cm^2$, $eV_{FB} \approx 1.65eV$) reveals that a larger interfacial layer capacitance $C_i$ characterizes the PLD counterpart. This indicates a higher interfacial quality for PLD-grown junctions.

*3.3 Mechanisms of RS suppression for the PLD-grown Schottky contacts*

Comparison of BEEM spectroscopy data for Au/NSTO junctions, grown respectively by PLD and by thermal evaporation, gives fruitful indications on the mechanisms that might underpin RS suppression in the PLD-grown SBDs. First of all, PLD metallization results into an average BEEM transmittance $\bar{R} \sim 3.8 \times 10^{-4} eV^{-5/2}$ across the Au/NSTO junctions (figure 6(a)), which is a factor $\sim 6$ higher than that measured for the evaporated counterpart i.e. $\bar{R} \sim 0.6 \times 10^{-4} eV^{-5/2}$ [36]. Since $\bar{R}$ is affected by the hot electrons tunneling probability across the thin interfacial layer (figure 7(a)), the increase of $\bar{R}$ for the PLD case might indicate a sizable decrease of the interfacial layer thickness $\delta_i$. This observation qualitatively agrees with the 30% larger capacitance $C_i \propto \delta_i^{-1}$ we estimated for the PLD-grown Au/NSTO devices ($C_i \approx 14\mu F/cm^2$) ) compared to the thermally-evaporated ones $(C_i \approx 11\mu F/cm^2)$ (figure 7(b)); this in fact corresponds to a decrease of the thickness $\delta_i/\epsilon_i$ from ~0.080nm to ~0.064nm (Supplementary section S12). Indeed, Mikheev *et al.*[19] established a strong connection between RS suppression and the reduction of $\delta_i/\epsilon_i$ (or the increase of $C_i$). According to a purely electronic picture, RS in M/NSTO junctions involves charge trapping/detrapping effects within the interface layer. A reversible modulation $\Delta\phi_{RS}$ of the Schottky barrier height takes place upon the voltage-driven variation of the trapped charge. As $C_i$ determines the voltage drop due to the trapped charge, the magnitude of $\Delta\phi_{RS} \propto C_i^{-1}$ directly scales with $\delta_i/\epsilon_i$ [19]. Hence, RS may vanish for sufficiently small $\delta_i/\epsilon_i$ values. Interestingly, we found that $\delta_i/\epsilon_i$ ~0.018nm of the PLD-grown Pt/NSTO SBDs (Supplementary Figure S12) fits the ~0.015nm value previously reported for RS-free, epitaxial Pt/NSTO(001) interfaces [19]. Another argument in favour of the reduction of $\delta_i$ for the PLD-grown devices relies on the small ideality factors, $n$~$1.1 - 1.6$. In fact, according to the MIS model, $n \approx 1 + C_d/C_i \propto 1 + \delta_i C_d$, (where $C_d$ is the depletion region capacitance) [19]. This is also in line with Shang *et al.* [29], who originally observed a strong weakening of RS effects in M/NSTO junctions governed by TE emission, with *n* as small as ~1.3. Finally, the reduction of $\delta_i$ qualitatively agrees with the improved interfacial homogeneity, with a barrier spread $\sigma$~50meV by a factor ~2 smaller than that of thermally-evaporated junctions [36].

Given the common interpretation of the low-permittivity interfacial layer as a strongly electron-depleted region, we speculate that its thinning might reflect, on one hand, local (intrinsic) chemical doping by oxygen vacancies generated by the PLD metallization. Both Au and Pt are high work function metals that are not expected to scavenge oxygen at the interface with $SrTiO_3$ at modest temperatures [60]. Nonetheless, NSTO surface reduction may occur at RT in the earliest stages of PLD metallization, because of the enhanced oxygen vacancy incorporation triggered either by the UV-radiation of the plasma plume or by the high kinetic energy of the impinging metal species [61,62]. The incorporation of interfacial oxygen vacancies explains the occurrence of $\phi_B$ values smaller than the Schottky-Mott limit, as previously discussed for M/NSTO interfaces prepared in a reduced state in conjunction with the concept of oxygen-vacancy-induced interface states [11–13]. On the other hand, resputtering and subsurface implantation by energetic metal ions [63] might contribute to make any carbon contamination layer thinner, and to promote intimate metal contact with NSTO [19,28].

The RS suppression of PLD-grown SBDs also relies on the fact that the environmental gas phase does not affect pristine cells functionality at RT over the typical working times. This is in sharp contrast to M/NSTO junctions prepared by thermal and e-beam evaporation, or by magnetron sputtering [21,24,26,27]. It is likely that the smooth and compact metal pads grown by PLD – thanks to the enhanced nucleation density and transient mobility of metal adatoms [64] – might offer a more efficient barrier against the field-assisted migration of ionic (charge-trapping) species, such as oxygen ions or water molecules, at the active interface. The above description allows to rationalize aging effects – particularly prominent for Au/NSTO Schottky contacts upon several-months-exposure to ambient air - in terms of a very slow re-oxidation of the interfacial oxygen vacancies which, as expected [11], leads to the increase of $\phi_B$.

## 4. CONCLUSIONS

Contrary to the common evidence that Schottky contacts prepared on NSTO by conventional RT metallization techniques display large interface-type RS effects, here we have documented the RS suppression in PLD-grown Au/ and Pt/NSTO junctions. Additionally, devices display electronic transport governed by TE and improved nanoscale spatial homogeneity compared to contacts fabricated by thermal evaporation. The RS suppression is ascribed to a higher interfacial quality, as attested by the small effective thickness $\delta_i/\epsilon_i \sim 0.02 - 0.06$nm of the (unintentional) low-permittivity interfacial layer. We speculate that in PLD, the UV-radiation of the plasma plume and the energetic impinging metal species either reduce the oxide surface layer – leading to $\phi_B \sim 0.70 - 0.80$ eV significantly below the Schottky-Mott limit – and do promote intimate metal contact. The poor sensitivity to environmental oxygen and moisture suggests that both interfacial chemical oxidation and electromigration of ionic (charge-trapping) species are greatly slowed down for the PLD-grown junctions. This fits the evidence that an appreciable increase of $\phi_B$ (by $\sim 0.2$ eV) takes place only upon several-months-exposure to ambient air. Overall, our study offers a viable strategy for the RT realization of robust, RS-free NSTO-based Schottky contacts.

**Acknowledgements**

This work was carried out within the framework of the project "RAISE - Robotics and AI for Socio-economic Empowerment" and has been supported by European Union - NextGenerationEU. However, the views and opinions expressed are those of the authors alone and do not necessarily reflect those of the European Union or the European Commission. Neither the European Union nor the European Commission can be held responsible for them.

**ORCID iDs**

R Buzio https://orcid.org/0000-0001-5632-5531

A Gerbi https://orcid.org/0000-0002-7323-2908

[54]  Goossens A S and Banerjee T 2023 Tunability of voltage pulse mediated memristive functionality by varying doping concentration in SrTiO3 *Appl. Phys. Lett.* **122**

[55]  Sullivan J P, Tung R T, Pinto M R and Graham W R 1991 Electron transport of inhomogeneous Schottky barriers: A numerical study *J. Appl. Phys.* **70** 7403–24

[56]  Polop C, Rosiepen C, Bleikamp S, Drese R, Mayer J, Dimyati A and Michely T 2007 The STM view of the initial stages of polycrystalline Ag film formation *New J. Phys.* **9** 74–74

[57]  Troadec C and Goh K E J 2010 Dual parameter ballistic electron emission spectroscopy analysis of inhomogeneous interfaces *J. Vac. Sci. Technol. B, Nanotechnol. Microelectron. Mater. Process. Meas. Phenom.* **28** C5F1-C5F4

[58]  Wrana D, Cieślik K, Belza W, Rodenbücher C, Szot K and Krok F 2019 Kelvin probe force microscopy work function characterization of transition metal oxide crystals under ongoing reduction and oxidation *Beilstein J. Nanotechnol.* **10** 1596–607

[59]  Aballe L, Matencio S, Foerster M, Barrena E, Sánchez F, Fontcuberta J and Ocal C 2015 Instability and Surface Potential Modulation of Self-Patterned (001)SrTiO$_3$ Surfaces *Chem. Mater.* **27** 6198–204

[60]  Posadas A B, Kormondy K J, Guo W, Ponath P, Geler-Kremer J, Hadamek T and Demkov A A 2017 Scavenging of oxygen from SrTiO 3 during oxide thin film deposition and the formation of interfacial 2DEG *J. Appl. Phys.* **105302** 1–10

[61]  Scullin M L, Ravichandran J, Yu C, Huijben M, Seidel J, Majumdar A and Ramesh R 2010 Pulsed laser deposition-induced reduction of SrTiO3 crystals *Acta Mater.* **58** 457–63

[62]  Hensling F V E, Keeble D J, Zhu J, Brose S, Xu C, Gunkel F, Danylyuk S, Nonnenmann S S, Egger W and Dittmann R 2018 UV radiation enhanced oxygen vacancy formation caused by the PLD plasma plume *Sci. Rep.* **8** 4–10

[63]  Perea A, Gonzalo J, Budtz-Jørgensen C, Epurescu G, Siegel J, Afonso C N and García-López J 2008 Quantification of self-sputtering and implantation during pulsed laser deposition of gold *J. Appl. Phys.* **104**

[64]  Shen J, Gai Z and Kirschner J 2004 Growth and magnetism of metallic thin films and multilayers by pulsed-laser deposition *Surf. Sci. Rep.* **52** 163–218

# Resistive switching suppression in metal/Nb:SrTiO$_3$ Schottky contacts prepared by room-temperature Pulsed Laser Deposition

R. Buzio[1,2*] and A. Gerbi[1,2]

[1]*CNR-SPIN Institute for Superconductivity, Innovative Materials and Devices, C.so Perrone 24, I-16152 Genova, Italy*

[2]*RAISE Ecosystem, Genova, Italy*

*Authors to whom any correspondence should be addressed.

E-mail: renato.buzio@spin.cnr.it

## SUPPLEMENTARY INFORMATION

## S1. C-V characteristics of low-doped Pt/NSTO SBDs

For pristine low-doped Pt/NSTO SBDs ($x_{Nb} = 0.01 \text{wt.} \%$) we obtained linear $C^{-2} vs\ V$ characteristics at RT, as expected for a field-independent dielectric constant $\epsilon_s$. The characteristics, reported in Figure S1a, showed a weak dependence on the modulation frequency $f_{CV}$ in the range 1KHz – 100KHz.

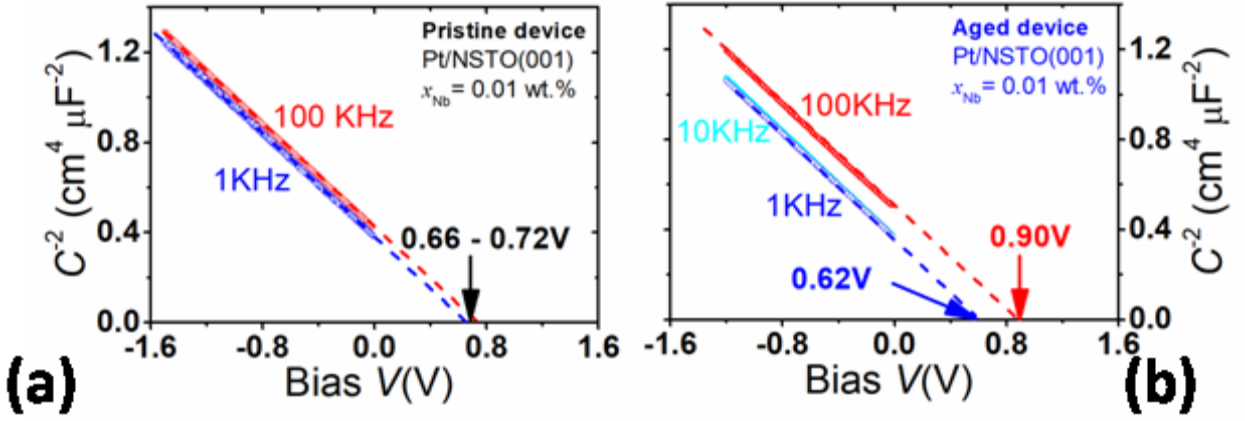

**Fig. S1.** (a) Room-temperature C-V characteristics for a representative pristine device: the horizontal intercepts 0.66-0.72eV depends weakly on the modulation frequency.(b) RT C-V characteristics for the same device in (a), upon aging in ambient air for 60 months. The frequency dependence of the junction capacitance suggests enhanced charge trapping at interface states.

Given the near-ideal RT transport properties, we interpolated the C-V measurements with the equation:

$$\frac{1}{C^2} = \frac{2}{\epsilon_0 \epsilon_r q N_D}(V_{bi} - V) \qquad (S1)$$

The fit parameters were $V_{bi}$ and $N_D$, and we assumed $\epsilon_r = 300$. We estimated the Schottky barrier height as $\phi_B^{C-V} = (eV_{bi} + \xi_F + k_B T)$, with $k_B T = 25\text{meV}$ and $\xi_F = 30\text{meV}$.[1] We obtained $N_D = 8.1 \times 10^{17} cm^{-3}$ and $\phi_B^{C-V} \approx 0.71 - 0.77\text{eV}$. The latter agrees with $\phi_B^{I-V} \cong 0.75\text{eV}$. Following the approach by Mikheev *et al.*[2], we also attempted to account for the presence of a low-permittivity (intrinsic) interfacial layer [2,3] by interpolating the C-V measurements with the equation:

$$\frac{1}{C^2} = \frac{2n^2}{\epsilon_0 \epsilon_r q N_D}(V_{bi} - \frac{V}{n}) \qquad (S2)$$

where $n > 1$ is the ideality index taken from the corresponding I-V characteristics. In this case, we estimated the interfacial layer capacitance as $C_i = \frac{n}{n-1}C$. With $n = 1.2$, we obtained $N_D = 9.7 - 9.8 \times 10^{17} cm^{-3}$, $\phi_B^{C-V} \approx 0.61 - 65\text{eV}$, $C_i \approx 9.2 - 10\mu F/cm^2$. The interfacial layer capacitance $C_i$ is of the same order of magnitude of that found for other NSTO-based SBDs [4] and corroborates the description of the system through the MIS model (see main text); $\phi_B^{C-V} < \phi_B^{I-V}$ however signals some overestimation of the interfacial layer effects.

In Figure S1b we also report RT $C^{-2} vs\ V$ characteristics for the same Pt/NSTO device after 60 months aging in air. There is now a prominent dependence on the modulation frequency $f_{CV}$, that corresponds (in the ideal-diode approximation of Eq. S1) to barrier height going from $\phi_B^{C-V} \approx 0.68\text{eV}$ ($f_{CV} = 1\text{KHz}$) up to $\phi_B^{C-V} \approx 0.90\text{eV}$ ($f_{CV} = 100\text{KHz}$). The estimated doping $N_D$ varies from $8.0 \times 10^{17} cm^{-3}$ ($f_{CV} = 1\text{KHz}$) to $8.1 \times 10^{17} cm^{-3}$ ($f_{CV} = 100\text{KHz}$). The frequency dependence of the junction capacitance, together with the situation $\phi_B^{C-V} \approx 0.90\text{eV} \gg \phi_B^{I-V} \approx 0.75\text{eV}$, were previously reported for Pt/NSTO diodes and ascribed to charge-trapping at interface states[5,6] and/or enhanced interfacial layer capacitance effects [2]. Spatial barrier inhomogeneity might also contribute [7]. Since Schafranek *et al.*[8] reported ~1.2eV for an oxidized Pt/NSTO interface, the barrier increase at $\phi_B^{C-V}(100\text{KHz}) \approx 0.90\text{eV}$ suggests the tendency to device reoxygenation upon aging. Notably, quasi-static I-V curves are not sensitive to such effects (see Figure 2d).

## S2. Macroscopic transport properties of highly-doped Pt/NSTO SBDs

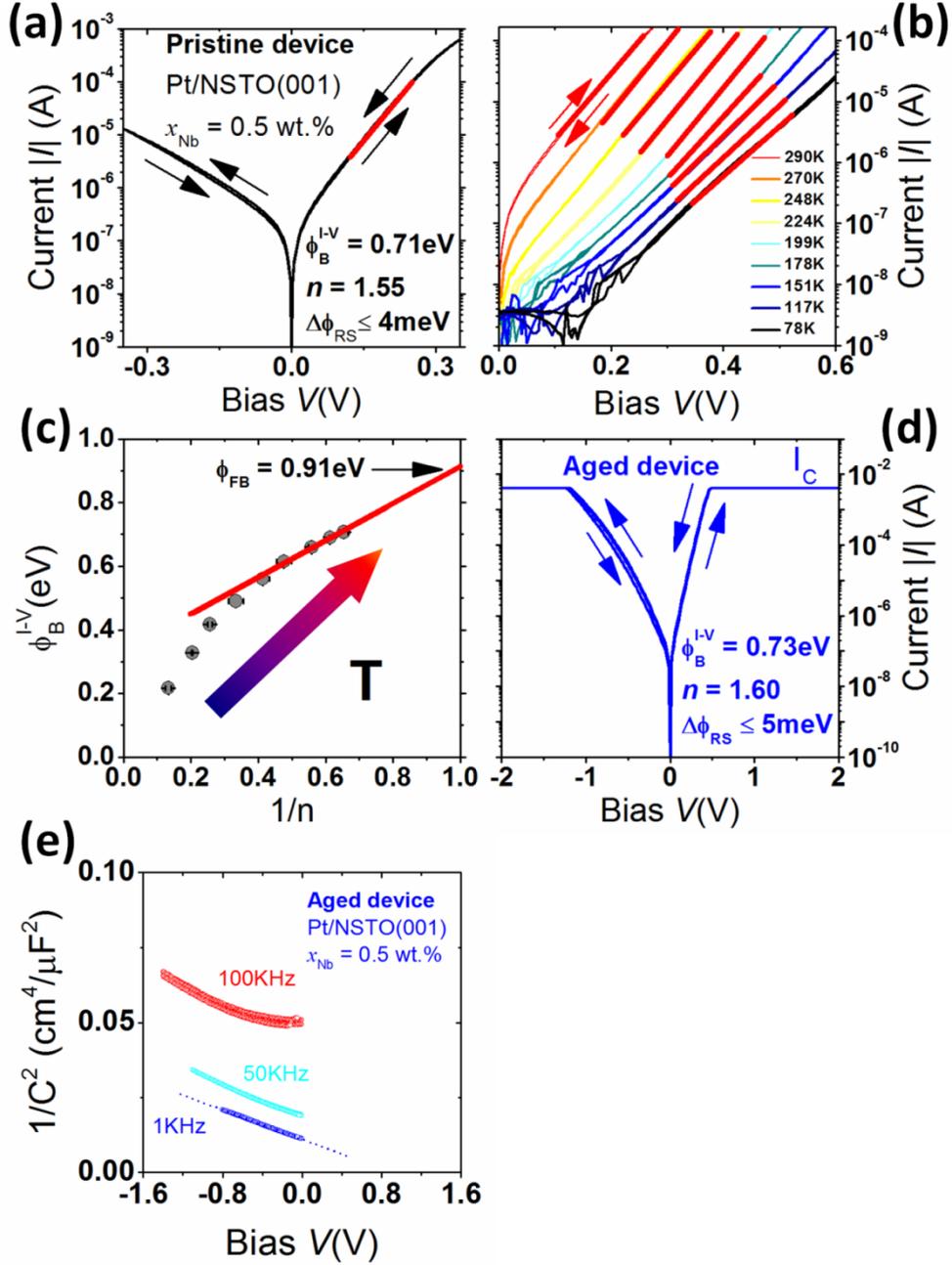

**Figure S2.** a) Highly reversible I–V curve of a pristine PLD-grown Pt/(001)NSTO SBD with $x_{Nb} = 0.5$ wt.%, measured at RT in UHV under cyclic polarization. Sweep directions are indicated by the arrows. Interpolation with the TE theory is shown (red curve). b) Temperature-dependent I–V curves (forward branch) for the device in (a), revealing highly-reversible characteristics at low temperatures. c) The $\phi_B^{I-V}$ vs $n^{-1}$ plot obtained from the I-V-T curves in b). Linear interpolation in the T range 175-290K (red curve) provides an estimate of the flat-band barrier height $\phi_{FB} = 0.91$eV consistent with the low-doping case. However, there is no linear correlation of experimental data over the whole temperature range. d) Highly reversible I–V curve measured at RT on the aged device (60 months aging in ambient air); cyclic polarization extends up to $\pm 2$V while keeping the maximum flowing current limited to $I_C = 5$mA. e) Room-temperature $C^{-2}$ vs $V$ characteristics for the device in (d): the field-dependent permittivity of NSTO is responsible for their curvature. For $f_{CV} = 1$KHz, the linear approximation (Eq.S2) gives $N_D = 2.2 \times 10^{20}$cm$^{-3}$, $\phi_B^{C-V} \approx 0.56$eV $< \phi_B^{I-V}$, $C_i \approx 27\mu$F/cm$^2$ (with $n = 1.60$, $\xi_F = -20$meV and $\epsilon_s = 85$)[2].

## S3. RS hysteresis under large polarization bias for Pt/NSTO SBDs

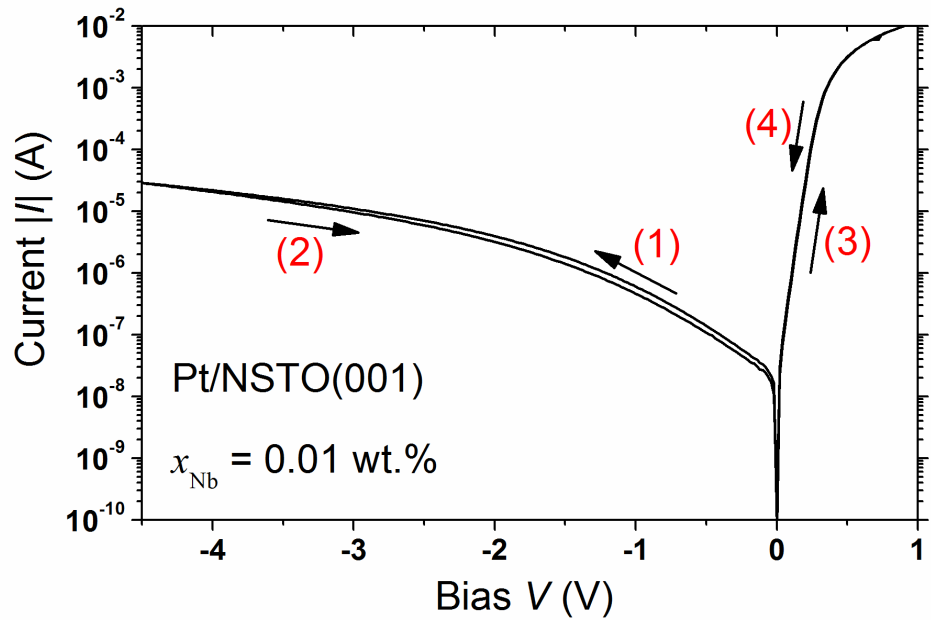

**Figure S3.** RT I-V characteristic acquired under cyclic polarization across the Pt/NSTO SBD of Figure 2(a). The sweep direction is highlighted by the numbered arrows 1 → 2 → 3 → 4. Despite the large polarization bias (-4.5V) in the backward branch, RS hysteresis is clearly suppressed in the forward branch.

## S4. Macroscopic transport properties of pristine and aged Au/NSTO SBDs

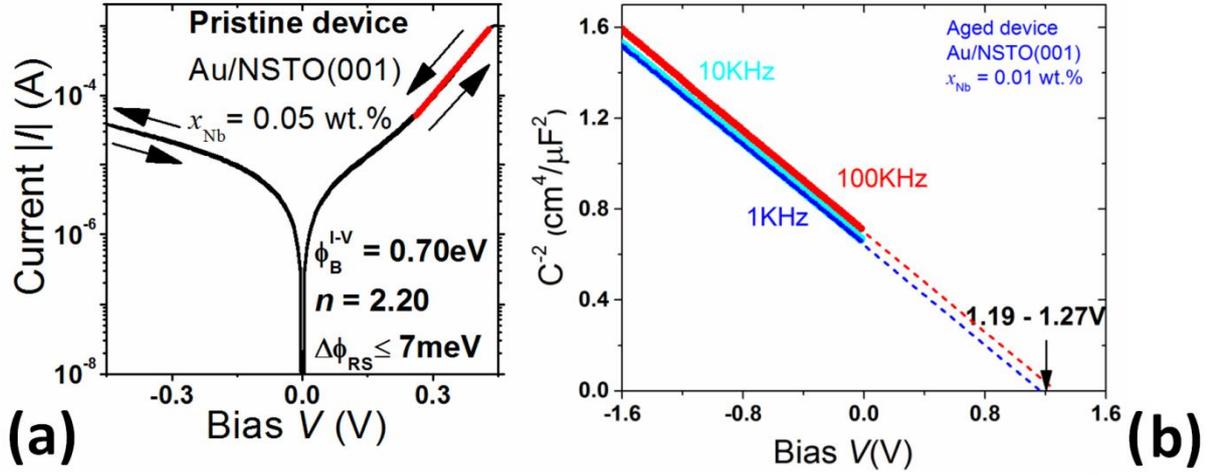

**Figure S4.** a) Highly reversible I–V curve of a PLD-grown Au/NSTO SBD with $x_{Nb} = 0.05$ wt.%, measured at RT in UHV under cyclic polarization. Sweep directions are indicated by the arrows. Interpolation with the TE emission is shown (red curve). b) Room-temperature $C^{-2} vs\ V$ characteristics for the aged device in Figure 3(d). Interpolation of data with the linear approximation (Eq.S2) gives $N_D \approx 1.1 \times 10^{18} \text{cm}^{-3}$, $\phi_B^{C-V} = 1.01 - 1.07\text{eV} \approx \phi_B^{I-V}$, $C_i \approx 6.3 \mu\text{F/cm}^2$ (with $n = 1.25$, $\xi_F = -30$meV and $\epsilon_s = 300$).

## S5. Morphology of PLD-grown ultrathin Au electrodes

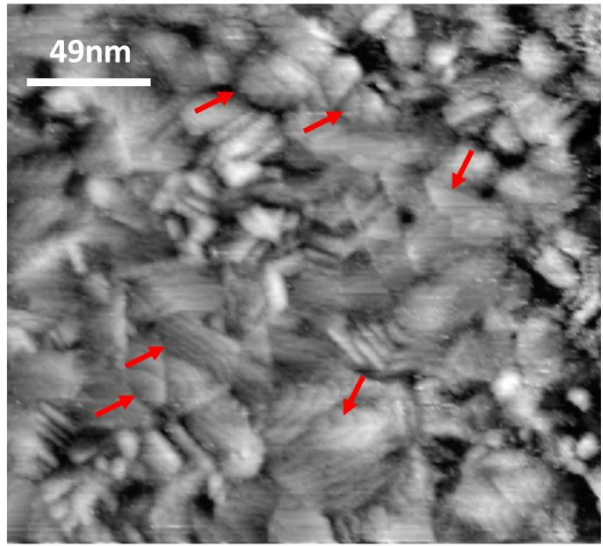

**Figure S5.** High-magnification STM topography of a PLD-grown Au electrode ($I_T = 20$nA, $V_T = -1.95$V, T = 80K). After wet-etching in Aqua Regia, the NSTO substrate ($x_{Nb} = 0.01$wt.%) was annealed at 1150°C in flowing $O_2$, and then exposed to PLD. Red arrows highlight atomically-smooth facets. A continuous Au film with atomically-smooth facets was observed also when the PLD growth took place on NSTO substrates annealed at 1000°C.

## S6. BEEM contrast at Au/TiO$_2$ and Au/SrO Schottky junctions

In the main text we associate the highly-transmitting BEEM domains (with $R \geq 4 \times 10^{-4} \text{eV}^{-5/2}$) to local Au/TiO$_2$ interfaces, and the low-transmitting regions ($R \leq 1.5 \times 10^{-4} \text{eV}^{-5/2}$) to local Au/SrO junctions. Different arguments support this conclusion. On one hand, by inspecting different surface spots of the same Au electrode by BEEM, we realized that the surface coverage of the 'low-transmitting' domains was always smaller than that of the 'highly-transmitting' ones (which corresponds to a fraction $20 - 40\%$ for the former and $60 - 80\%$ for the latter). Furthermore, we noticed that BEEM measurements carried out on SBDs prepared on NSTO substrate heated at 1000°C (rather than at 1150°C as in Figure 1) usually showed a negligible amount of 'low-transmitting' domains. In particular, BEEM maps did not show any spatial modulation of the BEEM signal (Figure S6 (a),(b)), having indeed a nearly isotropic autocorrelation function. Additionally, the hot electrons injection efficiency was as high as that recorded for the 'high-transmitting' domains on a mixed-terminated NSTO substrate (Figure S6(c)). This correlates with the smaller amount of SrO domains expected at the surface of NSTO after recrystallization at a lower temperature. Finally, according to background literature, SrO behaves as an insulating phase tunneled by hot electrons reaching the underneath TiO$_2$ layer. Such arguments led us to identify the 'low-transmitting' domains with the minority SrO-terminated domains.

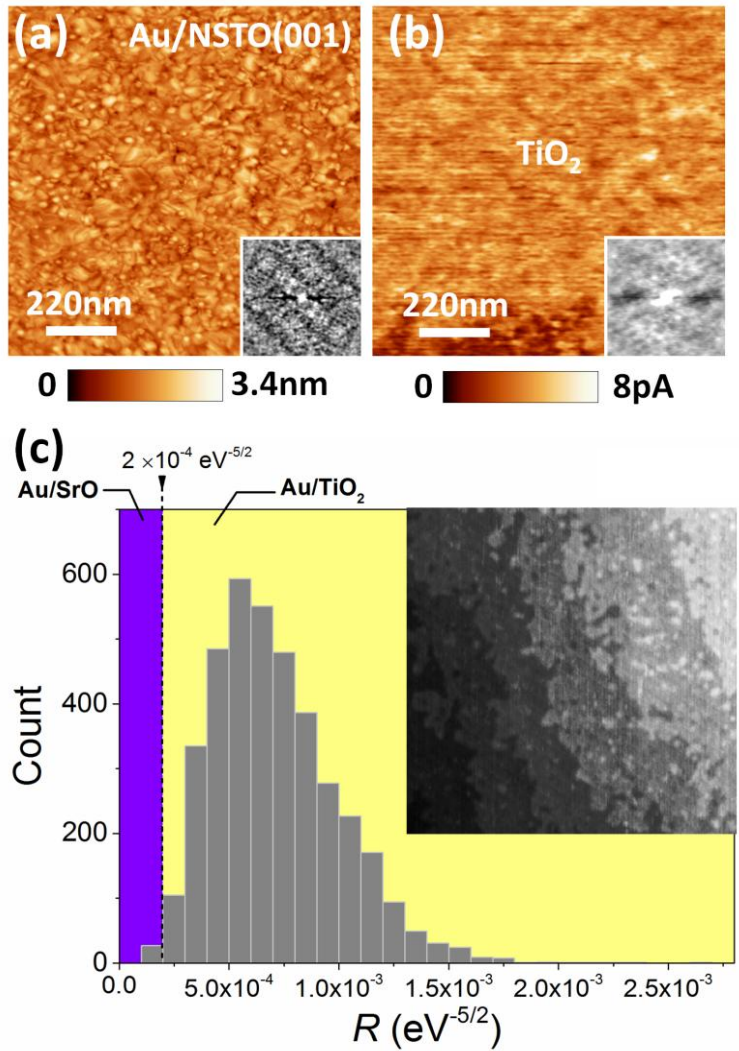

**Figure S6.** (a) STM topography and (b) BEEM map acquired simultaneously over a representative Au region of a PLD-grown Au/NSTO SBD ($I_T = 25nA$, $V_T = -1.60V$, RT). Here the NSTO substrate ($x_{Nb} = 0.01wt.\%$) was annealed at 1000°C in flowing $O_2$ before exposure to PLD. The inset in (a) is the autocorrelation function of the topography. (b) BEEM map - acquired simultaneously to the map in (a) - does not reveal any spatial modulation associated to a mixed termination of the underlying substrate. The inset in (b) is the nearly isotropic autocorrelation function of the BEEM map. (c) Histogram of the ballistic transmittance from a set of about 3800 BEEM spectra acquired on the SBD in (a),(b): most of the ballistic transmittance values are above $\sim 2 \times 10^{-4} eV^{-5/2}$ i.e. comparable to BEEM spectra acquired at the Au/TiO$_2$ interface. The inset is an AFM topography showing the single-step terraced structure of the underlying NSTO substrate.

## S7. BEEM maps on mixed terminated NSTO: the Au/NSTO case

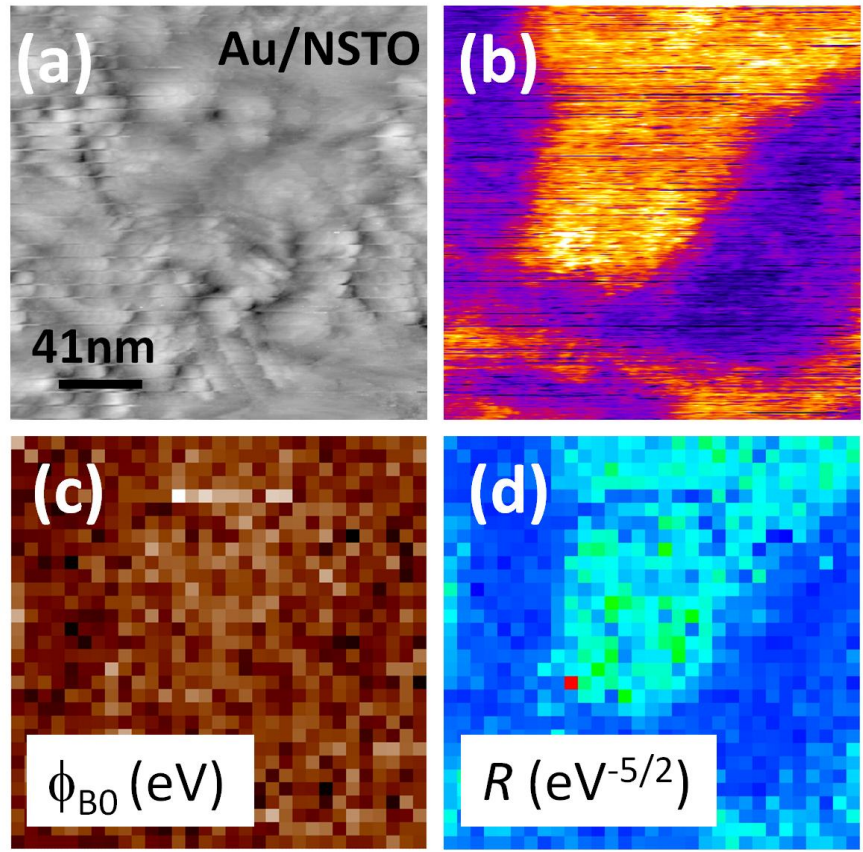

**Figure S7.** (a) STM topography and (b) BEEM map acquired simultaneously over a representative Au region of the Au/NSTO SBD considered in Figure 6(a). (c,d) spatially-resolved maps for $\phi_{B0}$ and $R$, extracted from analysis of a grid of 31 × 31 spectra acquired over the region in (b). Note that the mixed termination affects mostly the $R$ map.

## S8. Average BEEM spectra on mixed terminated NSTO: the Au/NSTO case

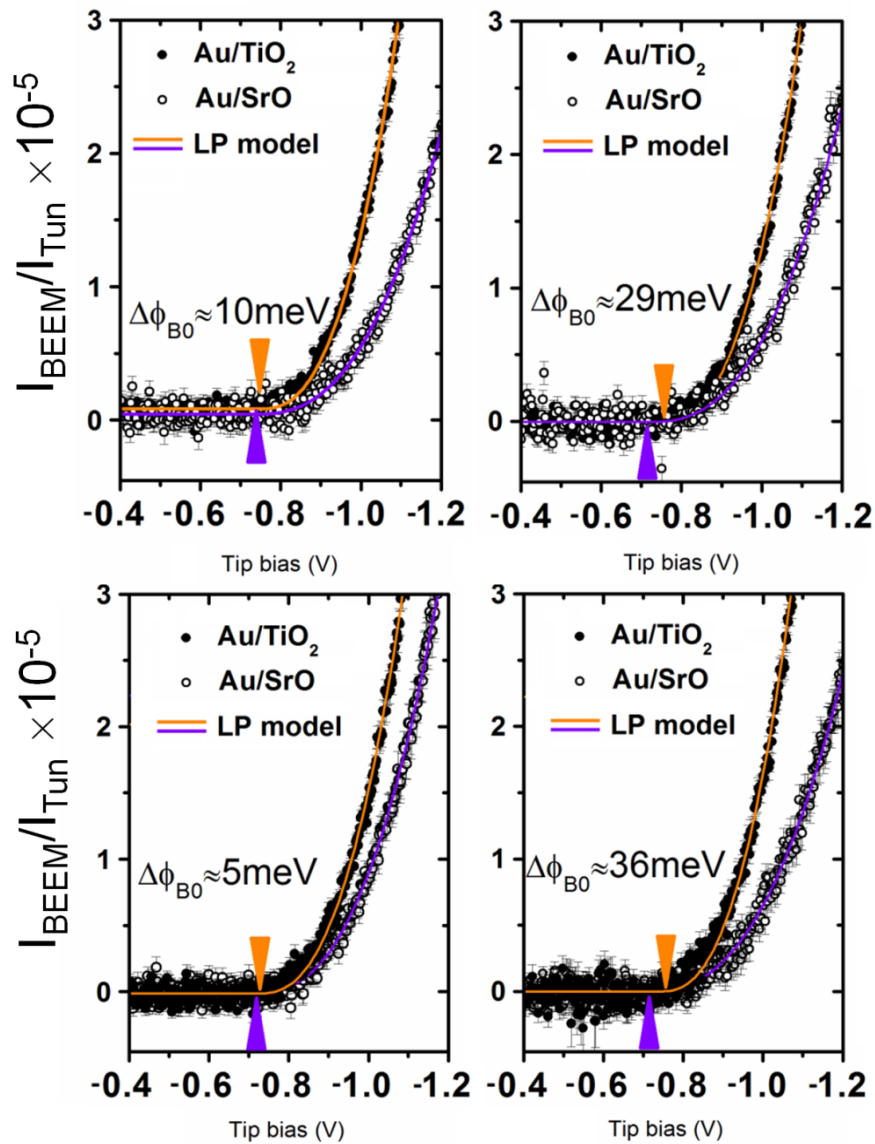

**Figure S8.** Average BEEM spectra for a PLD-grown Au/NSTO SBD, acquired on selected regions associated to different surface terminations of NSTO ($x_{NB} = 0.01$ wt. %; T = 80K).

## S9. Noise contributions to BEEM current

We estimated the contribution of the signal-to-noise (S/N) ratio to the statistical spread by the method of C. Troadec and K.E.J. Goh [9]. We simulated BEEM spectra using the power law of the LP model, with barrier height and transmission factor comparable with the measured ones (Figure 6). Furthermore, a Gaussian white noise with standard deviation 120fA (equal to the RMS current fluctuation of the BEEM signal with a SBD connected at the input, and an injected tunnelling current $I_T \sim 20$nA) was added to the power law. Then, each spectrum was fitted with the LP model under the same conditions used for experimental spectra (interpolation range $0.4V - 1.2V$). Finally, the statistical spread of the quantity $\phi_{B0} = e|V_{th,SB}|$ was quantified. Figure S9 (left panel) shows the results for 1000 simulated spectra, corresponding to the T=80K situation with mean barrier height 0.8eV and mean attenuation factor $R \sim 5 \times 10^{-4} \text{eV}^{-5/2}$. The statistical spread of the simulated distribution gives a standard deviation of 8.5meV (FWHM = 17meV). Likewise, in Figure S9 (right panel) we consider the spread of the simulated distribution corresponding to the smaller barrier height 0.7eV with attenuation factor $R \sim 1 \times 10^{-4} \text{eV}^{-5/2}$ (roughly corresponding to the low-barriers tail of the distribution in Figure 6(b)). In such case the statistical spread of the simulated distribution corresponds to a standard deviation of 28meV (FWHM = 56meV).

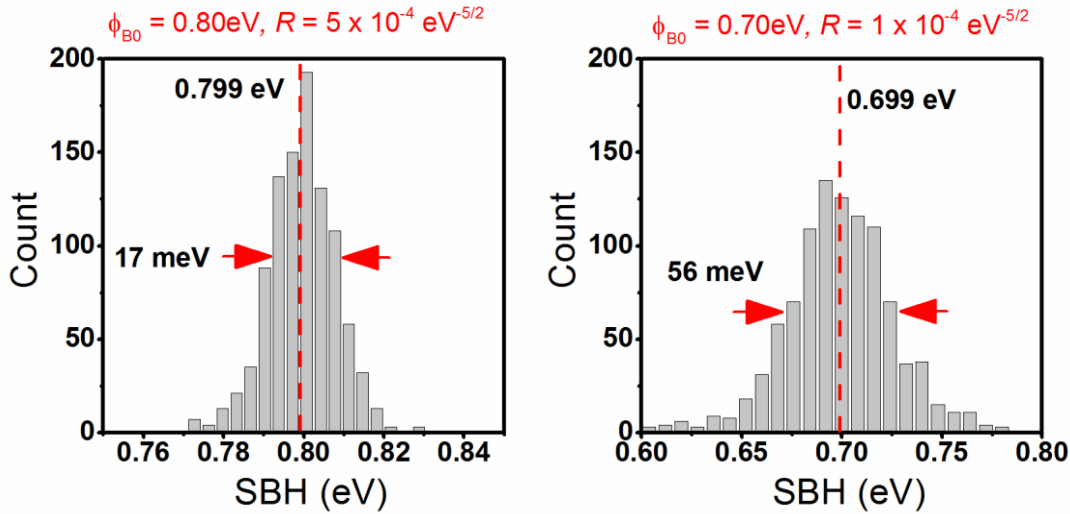

**Figure S9.** Distributions of barriers for simulated BEEM spectra, generated according to the $\phi_{B0}$ and $R$ values indicated in each panel (parameters in red). A Gaussian white noise with standard deviation 120fA was superimposed on each spectrum before the LP fit.

For simplicity, we assumed an overall spread contribution from the S/N ratio of about $\sigma_{simu} = \sqrt{8.5^2 + 28^2}$ meV $\approx 30$ meV. One thus concludes that intrinsic interfacial inhomogeneity does emerge above the experimental uncertainty. The actual variation of the Schottky barrier height can be deduced from the measured distributions approximately as $\sqrt{\sigma_{exp}^2 - \sigma_{simu}^2}$ [10]. This choice results in the standard deviations $\sim 50 - 55$ meV reported in the main text.

## S10. BEEM maps on mixed terminated NSTO: the Pt/NSTO case

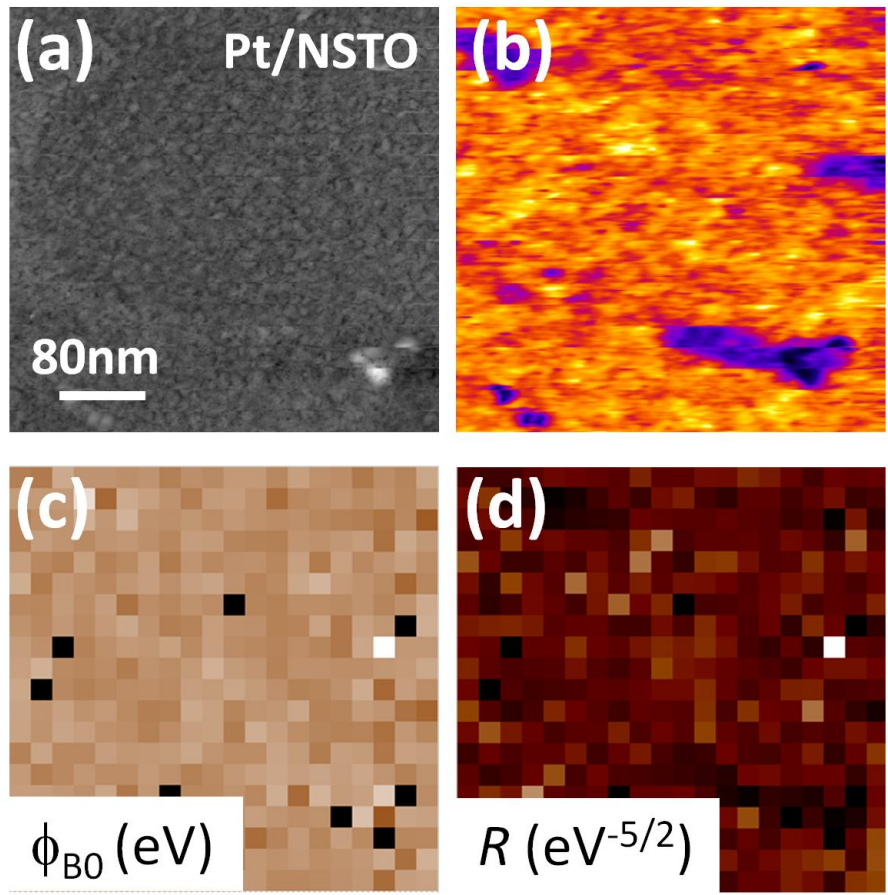

**Figure S10** (a) STM topography and (b) BEEM map acquired simultaneously over a representative Pt region of the Pt/NSTO SBD in Figure 6(c). (c,d) spatially-resolved maps for $\phi_{B0}$ and $R$, extracted from analysis of a grid of $20 \times 20$ spectra acquired over the region in (b). Note that the mixed termination affects mostly the $R$ map. Black pixels in (c),(d) correspond to surface spots that were intentionally discarded from BEEM spectra analysis due to instrumental artifacts.

## S11. Average BEEM spectra on mixed terminated NSTO: the Pt/NSTO case

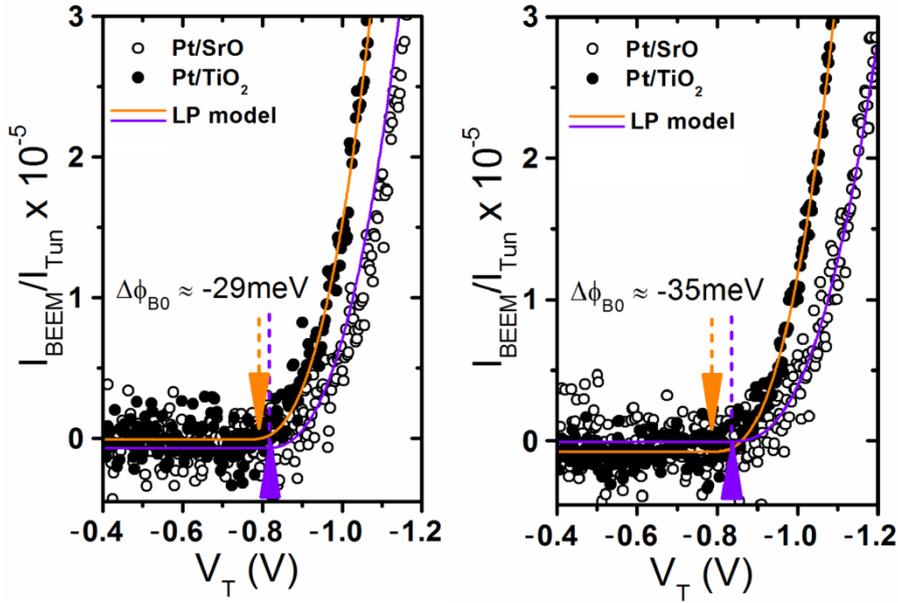

**Figure S11.** Average BEEM spectra for a PLD Pt/NSTO SBD, acquired on selected regions associated to different surface terminations of NSTO ($x_{NB} = 0.01$ wt.%; T=295K).

## S12. Electrostatic MIS model for the M/NSTO junction

For a detailed description of the MIS model see the Supplementary Information Section S6 in [11]. Briefly, we have calculated the zero-bias Schottky barrier height $\phi_{B0}$ at each temperature T as $\phi_{B0}(T) = |\psi_S(T) + \xi_F|$, where $\xi_F$ is the Fermi level measured from the bottom of the conduction band of NSTO (in the neutral semiconductor region, $\xi_F > 0$ for a non-degenerate semiconductor) and $\psi_S(T)$ is the surface potential at the insulator/NSTO interface. According to the notation used in [12], $\psi_S(T) < 0$. The value of $\psi_S(T)$ was obtained by solving numerically the equation [12]:

$$\psi_S + eV_{FB} - \frac{b(T)\epsilon_0}{C_i}\operatorname{arccosh}\left(1 - \frac{eN_D}{\sqrt{a(T)}b(T)\epsilon_0}\psi_S\right) = 0 \tag{S3}$$

with $eV_{FB}$ flat band voltage, $C_i^{-1} = \delta_i/\epsilon_i\epsilon_0$ inverse areal capacitance of the interfacial layer, $N_D$ carriers density, $b(T)$ and $a(T)$ material parameters of the temperature- and field-dependent permittivity $\epsilon_s(E,T) = b(T)/\sqrt{a(T) + E^2}$ of NSTO. The parameters $N_D = 0.9 \times 10^{18} cm^{-3}$ and $\xi_F = 40 meV$ were kept fixed. The remaining parameters, $eV_{FB}$ and $C_i$, were iteratively adjusted to minimize the discrepancy between experimental $\overline{\phi}_{B0}$ vs T curves (Figure 7(b)) and numerical predictions from the MIS model.

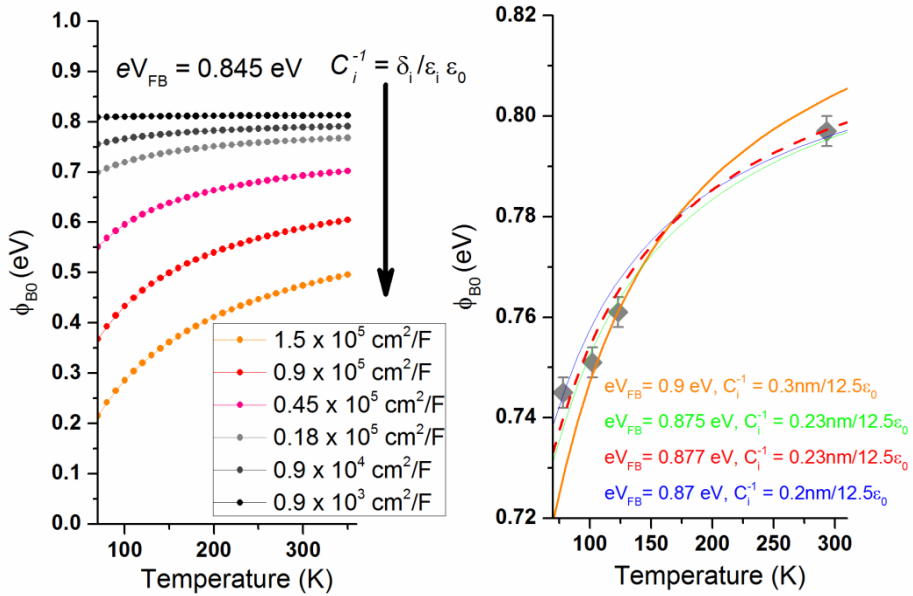

**Figure S12.** (Left panel) Predictions of the MIS model for a constant flat-band barrier height $eV_{FB} = 0.845 eV$, and increasing values of $C_i^{-1}$ (from top to bottom). (Right panel) Comparison of experimental BEEM data for the PLD-grown Pt/NSTO SBD with predictions of the MIS model, for different pairs of the parameters ($eV_{FB}$, $C_i^{-1}$).

Below we resume the parameters used to interpolate BEEM data of Figure 7(b).

|  | Au/NSTO Thermal evap. (from [11]) | Au/NSTO PLD (this study) | Pt/NSTO PLD (this study) |
|---|---|---|---|
| $N_D (cm^{-3})$ | $0.9 \times 10^{18}$ | $0.9 \times 10^{18}$ | $0.9 \times 10^{18}$ |
| $\xi_F (eV)$ | 0.04 | 0.04 | 0.04 |
| $eV_{FB} (eV)$ | **1.58** | **1.13** | **0.88** |
| $C_i^{-1} (\times 10^5 cm^2/F)$ | **0.9** | **0.7** | **0.2** |
| $\delta_i/\epsilon_i$ (nm) | **0.080** | **0.064** | **0.018** |